\documentclass[11pt]{article}

\usepackage[T1]{fontenc}
\usepackage[utf8]{inputenc}
\usepackage[margin=1in]{geometry}
\usepackage{amsmath,amssymb,amsthm,mathtools}
\usepackage{enumitem}
\usepackage{booktabs}
\usepackage{xcolor}
\usepackage{tikz}
\usetikzlibrary{arrows.meta,positioning}
\usepackage{float}
\floatstyle{ruled}
\newfloat{algorithm}{t}{lop}
\floatname{algorithm}{Algorithm}
\usepackage{hyperref}
\usepackage{cleveref}
\usepackage{authblk}

\hypersetup{
    colorlinks=true,
    linkcolor=blue!70!black,
    citecolor=blue!70!black,
    urlcolor=blue!70!black
}

\newtheorem{theorem}{Theorem}[section]
\newtheorem{lemma}[theorem]{Lemma}
\newtheorem{proposition}[theorem]{Proposition}
\newtheorem{corollary}[theorem]{Corollary}
\theoremstyle{definition}
\newtheorem{definition}[theorem]{Definition}
\newtheorem{remark}[theorem]{Remark}

\newcommand{\E}{\mathbb{E}}

\newcommand{\R}{\mathbb{R}}
\newcommand{\cC}{\mathcal{C}}
\newcommand{\cD}{\mathcal{D}}
\newcommand{\cU}{\mathcal{U}}
\newcommand{\cI}{\mathcal{I}}
\newcommand{\OPT}{\mathrm{OPT}}
\newcommand{\ALG}{\mathrm{ALG}}
\newcommand{\FB}{\textup{\textsc{Fair-Bias}}}
\newcommand{\EFB}{\textup{\textsc{Empirical-Fair-Bias}}}
\newcommand{\MST}{\mathrm{MST}}
\newcommand{\wt}{\mathrm{wt}}

\newcommand{\new}[1]{{#1}}
\newcommand{\old}[1]{}

\title{$O(1)$-Competitive Stochastic Online Metric Matching \\ in General Metrics} 

\author[1]{Jittat Fakcharoenphol}
\author[2]{Pasin Manurangsi}
\author[3]{Phanu Vajanopath\thanks{Supported by Polish National Science Centre grant 2020/39/B/ST6/01641.}}
\affil[1]{Kasetsart University, Thailand}
\affil[2]{Google Research, Thailand}
\affil[3]{University of Wroc{\l}aw, Poland}
\date{}

\begin{document}

\maketitle

\begin{abstract}
We study stochastic online metric matching, in which $n$ requests drawn independently from a common distribution must be matched immediately and irrevocably to $n$ servers in a metric space. Gupta, Guruganesh, Peng, and Wajc [ICALP'19] introduced the \textsc{Fair-Bias} algorithm and proved an $O((\log\log\log n)^2)$ competitive ratio on general metrics. We prove that \textsc{Fair-Bias} is $O(1)$-competitive on every finite metric space when the arrival distribution is known, resolving their open question.

Their analysis combines a ``balls-and-bins'' argument when many free servers remain with a tree-embedding argument when few remain. We replace this division into two regimes with a unified clustered transport analysis. We partition the servers into clusters along a minimum spanning tree tour (as in the twice-around-the-tree algorithm for metric TSP) and extend the balls-and-bins argument from individual servers to clusters to bound the intercluster cost. To bound the intracluster cost, we prove a sharp inequality showing that the minimum spanning tree weight is at most $e$ times the expected offline optimum under arrivals uniform over the $n$ servers.

For unknown arrival distributions, we introduce \textsc{Empirical-Fair-Bias}, a $(4\rho+1)$-competitive algorithm, where $\rho$ is the competitive ratio of \textsc{Fair-Bias} under uniform arrivals on the same metric space. Consequently, our algorithm is $O(1)$-competitive on general metrics. Concurrently and independently, Harada [arXiv'26] gives a different algorithm that also yields $O(\rho)$ competitiveness.
\end{abstract}

\emph{We used AI models at several stages of this work. In particular, both of the main results were originally discovered by Gemini 4 Carbon. These were then simplified further by the authors, with help from AI models. (See ``AI Disclosure'' for more detail.)}


\section{Introduction}
\label{sec:intro}
We study the following online metric matching problem.
Let $(X, d)$ be a finite metric space. A multiset of $n$ \emph{servers} $S = \{s_1, \dots, s_n\} \subseteq X$ is placed in $X$ before the game begins. 

Over $n$ time steps $t = 1, 2, \dots, n$, a sequence of $n$ \emph{requests} $r_1, r_2, \dots, r_n \in X$ arrives one by one. When request $r_t$ arrives at step $t$, there are $k \;:=\; n - t + 1$
unmatched (\emph{free}) servers remaining, which we denote by $S_k \subseteq S$ (initially $S_n = S$). The online algorithm must immediately and irrevocably choose one free server $s_t \in S_k$ to serve $r_t$. This choice incurs a matching cost of $d(s_t, r_t)$, and the server $s_t$ is removed from the free pool ($S_{k-1} \gets S_k \setminus \{s_t\}$). After step $n$, every request has been matched to a distinct server, producing a one-to-one matching between $R_n := (r_1, \dots, r_n)$ and $S$ of total cost $\sum_{t=1}^n d(s_t, r_t)$.

\paragraph{The Offline Benchmark and Competitive Ratio.}
If an algorithm knew the entire request sequence $R_n = (r_1, \dots, r_n)$ in advance, it could compute a \emph{minimum-cost perfect matching}: a bijection $\sigma : [n] \to [n]$ 
minimizing the total distance
\[
    \OPT(R_n, S) \;:=\; \min_{\sigma} \sum_{i=1}^n d(r_i, s_{\sigma(i)}).
\]
In the classic \emph{adversarial model}, where an adversary chooses the request sequence $r_1, \dots, r_n$ in the worst case, the optimal competitive ratio of \emph{deterministic} online algorithms is $2n-1$ on general metrics~\cite{KP93,KMV94} (whereas on a line, deterministic algorithms achieve $O(\log n)$~\cite{Rag18}). \emph{Randomized} online algorithms (against an oblivious adversary, who fixes $R_n$ without seeing the algorithm's random choices) improve the worst-case competitive ratio to $O(\log^2 n)$ on general metrics~\cite{BBGN14} and $O(\log n)$ on doubling metrics~\cite{GL12}---but every randomized online algorithm still requires $\Omega(\log n)$ on general metrics (even on a star metric)~\cite{MNP06} and $\Omega(\sqrt{\log n})$ on a line~\cite{PS21}. Even in the \emph{random-order model}---where an adversary chooses the multiset of $n$ requests upfront, and they arrive in a uniformly random order---Raghvendra~\cite{Rag16} showed that the optimal competitive ratio is $\Theta(\log n)$, so no online algorithm can achieve an $O(1)$ competitive ratio.

\paragraph{The Stochastic (i.i.d.) Model.}
To move beyond worst-case inputs, Gupta, Guruganesh, Peng, and Wajc~\cite{GGPW19} introduced the \emph{stochastic online metric matching} model: the requests $r_1, \dots, r_n$ are drawn \emph{independently and identically distributed (i.i.d.)} from an underlying probability distribution $\cD$ over $X$ (written $R_n \sim \cD^n$). For an online algorithm $\ALG$, let $\ALG(\cD)$ and $\OPT_{\cD} := \OPT(R_n, S)$ denote the random variables equal to the algorithm's total matching cost and the offline optimal matching cost on the random request sequence $R_n \sim \cD^n$, respectively. We compare the algorithm's expected total cost $\E[\ALG(\cD)]$ against the expected offline optimum
\[
    \E[\OPT_{\cD}] \;=\; \E_{R_n \sim \cD^n}\bigl[\OPT(R_n, S)\bigr].
\]
We say that an online algorithm is \emph{$C$-competitive} if $\E[\ALG(\cD)] \le C \cdot \E[\OPT_{\cD}]$.

When the distribution $\cD$ is \textbf{known} to the algorithm upfront, a natural algorithm called \FB{} (Algorithm~\ref{alg:fb} in Section~\ref{sec:overview}) was proposed in~\cite{GGPW19}: at step $t$ (with $k = n - t + 1$ free servers $S_k$), compute a minimum-cost way to transport between the uniform distribution $\cU_{S_k}$ on the remaining free servers (which places weight $1/k$ on each server in $S_k$) and $\cD$, and use this transport plan to match $r_t$ to $S_k$. \FB{} was shown in~\cite{GGPW19} to be $9$-competitive on line and tree metrics and $O((\log \log \log n)^2)$-competitive on general $n$-point metric spaces. Subsequent work achieved $O(1)$-competitiveness in Euclidean spaces $[0, 1]^d$ under additional structural assumptions: Kanoria~\cite{Kan22} analyzed dynamic spatial matching when both servers and requests are drawn i.i.d.\ in $[0, 1]^d$; Saberi, Yang, and Yu~\cite{SYY24} (and the journal version by Yang and Yu~\cite{YY26}) reduced adversarial server placements with known i.i.d.\ requests to i.i.d.\ server-request pairs, obtaining $O(1)$-competitiveness on $[0, 1]^d$ ($d \ge 3$) for smooth distributions; and for unknown (even non-identical) smoothed distributions on $[0, 1]^d$ ($d \ne 2$), Li, Vitercik, and Yang~\cite{LVY26} obtained $O(1)$-competitiveness assuming access to a single prior sample from each request's distribution. \old{For general finite metric spaces and arbitrary arrival distributions $\cD$ (without prior samples), however, two central open questions remained~\cite{GGPW19,YY26}:}

When $\cD$ is \textbf{unknown} upfront (the \emph{unknown i.i.d.\ model}), the problem is obviously at least as hard as with a known distribution and is at most as hard as in the random-order model, since conditioned on their multiset, i.i.d.\ requests arrive in a uniformly random order. Hence the optimal competitive ratios (CR) of the four arrival models satisfy
\[
    \mathrm{CR}(\textsc{Adversarial}) \;\ge\; \mathrm{CR}(\textsc{RandomOrder}) \;\ge\; \mathrm{CR}(\textsc{UnknownIID}) \;\ge\; \mathrm{CR}(\textsc{KnownIID}).
\]
For general finite metric spaces, two open questions in \cite{GGPW19} are the following:
\begin{enumerate}[leftmargin=2.5em,label=\textbf{(Q\arabic*)}]
    \item \label{q:o1} \emph{Does there exist an $O(1)$-competitive online algorithm on \textbf{every} finite metric space when the arrival distribution $\cD$ is known?}
    \item \label{q:unknown} \emph{\old{Can an online algorithm achieve an $O(1)$ competitive ratio when the arrival distribution $\cD$ is \textbf{unknown} upfront (the Unknown i.i.d.\ model)?}\new{How do the competitive ratios of these arrival models relate? In particular, is the unknown i.i.d.\ model closer to the known i.i.d.\ model or to the random-order model?}}
\end{enumerate}

\subsection{Main Results}

\old{We answer both~\ref{q:o1} and~\ref{q:unknown} affirmatively.} \new{We answer~\ref{q:o1} affirmatively and, as a consequence, settle~\ref{q:unknown} on general metrics up to constant factors.} For~\ref{q:o1}, we show that the aforementioned \FB{} algorithm of~\cite{GGPW19} is $O(1)$-competitive on every finite metric space.

\begin{theorem}[$O(1)$-Competitiveness on General Metric Spaces]
\label{thm:main-o1}
For every finite metric space $(X, d)$, every multiset of $n \ge 1$ servers $S \subseteq X$, and every known arrival distribution $\cD$ over $X$, the expected cost of \FB{} satisfies
\[
    \E[\FB(\cD)] \;\le\; O(1) \cdot \E[\OPT_{\cD}].
\]
\end{theorem}

A key ingredient, which may be of independent interest, is a (sharp) bound that the minimum spanning tree weight is at most $e$ times $\E[\OPT_{\cU_S}]$ when $\cU_S$ is the uniform distribution over $S$ (\Cref{thm:mst-sharp-bound}).



For~\ref{q:unknown}, we design a simple, parameter-free online algorithm called \EFB{} (Algorithm~\ref{alg:efb} in Section~\ref{sec:unknown}) and prove a universal reduction relating its cost directly to \FB{} on the uniform distribution $\cU_S$ over $S$ (which places weight $1/|S|$ on each server). Specifically, if \FB{} is $\rho$-competitive on $\cU_S$, then \EFB{} is $(4\rho + 1)$-competitive on every unknown arrival distribution $\cD$. This is stated more formally below.

\begin{theorem}[Reduction for Unknown Arrival Distributions]
\label{thm:main-unknown}
For every finite metric space $(X, d)$, every server multiset $S \subseteq X$ of size $n \ge 1$, and every \emph{unknown} arrival distribution $\cD$ over $X$, 
if $\E[\FB(\cU_S)]\le\rho\,\E[\OPT_{\cU_S}]$, then the expected cost of \EFB{} satisfies \[
    \E[\EFB(\cD)] \;\le\; (4\rho+1)\, \E[\OPT_{\cD}].
\]
\end{theorem}


Combining Theorem~\ref{thm:main-unknown} and Theorem~\ref{thm:main-o1} immediately gives $O(1)$-competitiveness for unknown distributions on all metrics.

\begin{corollary}[$O(1)$-Competitiveness for Unknown i.i.d.\ Arrivals]
\label{cor:unknown-o1}
For every finite metric space $(X, d)$, every server multiset $S \subseteq X$ of size $n \ge 1$, and every unknown arrival distribution $\cD$ over $X$, \EFB{} achieves an $O(1)$ competitive ratio:
\[
    \E[\EFB(\cD)] \;\le\; O(1) \cdot \E[\OPT_{\cD}].
\]
\end{corollary}

In particular, \Cref{cor:unknown-o1} shows that the unknown i.i.d.\ model admits an $O(1)$-competitive algorithm on all metric spaces, strictly separating it from the Random-Order model where $\Omega(\log n)$ is required~\cite{Rag16}. \new{Together with \Cref{thm:main-o1}, this answers~\ref{q:unknown} on general metrics: the optimal competitive ratios of the known and the unknown i.i.d.\ models are both $\Theta(1)$, whereas the random-order model requires $\Theta(\log n)$ (\Cref{tab:models}). Whether the adversarial (randomized) and random-order models are separated on general metrics remains open.}

\begin{table}[t]
\centering
\caption{\new{Optimal competitive ratios of the arrival models on general metrics.}}
\label{tab:models}
\begin{tabular}{lll}
\toprule
\new{Arrival Model} & \new{Competitive Ratio} & \new{Source} \\
\midrule
\new{Adversarial (Deterministic)} & \new{$2n-1$} & \new{\cite{KP93,KMV94}} \\
\new{Adversarial (Randomized)} & \new{$O(\log^2 n)$ and $\Omega(\log n)$} & \new{\cite{BBGN14,MNP06}} \\
\new{Random Order} & \new{$\Theta(\log n)$} & \new{\cite{Rag16}} \\
\new{Unknown IID} & \new{$\Theta(1)$} & \new{Corollary~\ref{cor:unknown-o1}} \\
\new{Known IID} & \new{$\Theta(1)$} & \new{Theorem~\ref{thm:main-o1}} \\
\bottomrule
\end{tabular}
\end{table}

\paragraph{\new{Concurrent work.}}
Concurrently and independently, Harada~\cite{Har26} studied~\ref{q:unknown} and gave an algorithm for unknown i.i.d.\ arrivals that is $O(\rho)$-competitive whenever \FB{}$(\cU_S)$ is $\rho$-competitive, similar to our \Cref{thm:main-unknown}. Thus, their result also gives an $O(1)$ competitive ratio on general metrics (\Cref{cor:unknown-o1}) when combined with our known i.i.d.\ result (\Cref{thm:main-o1}).


\subsection{Further Related Work}
\label{subsec:related-work}

For adversarial arrivals, Nayyar and Raghvendra~\cite{NR17} bounded the competitive ratio of Raghvendra's deterministic robust matching algorithm by $O(\mu(S)\log^2 n)$, where $\mu(S)$ is the maximum ratio of shortest TSP tour length to diameter over server subsets of positive diameter. They also proved an $\Omega(\mu(S))$ lower bound for every deterministic algorithm on the same server placement.

Against an oblivious adversary, \cite{CP08} obtained an $o(\log^3 n)$ bound. On $\Delta$-ary hierarchically separated trees, \cite{GRV26} obtained an $O(\log\Delta\log\log\Delta)$ ratio for large $\Delta$, nearly matching the $\Omega(\log\Delta)$ lower bound. 
Under random-order arrivals, greedy matching has been studied on general metrics~\cite{GK19} and on the uniform metric~\cite{DSX22}.


On the line, deterministic upper bounds improved from the work-function approach~\cite{KN04} and the first sublinear bound~\cite{ABNPS19} to $O(\log^2 n)$~\cite{NR17}; deterministic lower bounds exceed $9$~\cite{FHK05} and are stronger for restricted classes of algorithms~\cite{AFT18}. With uniform servers and requests on an interval, greedy is constant-competitive~\cite{BFP23}.

Beyond irrevocable matching, \cite{GKS20} obtained an $O(\log n)$ ratio on general metrics with $O(\log n)$ recourse per request and a constant ratio on the line with $O(\log n)$ amortized recourse, the latter also obtained independently by \cite{MN25}. Predictions and advice have also been studied~\cite{ACEPS23,SV26,CCM25,CN24}.

\subsection{Organization of the Paper}
\label{subsec:organization}
\Cref{sec:overview} gives an overview of both proofs. \Cref{sec:prelim} introduces our notation and the tools from optimal transport and from the analysis of \FB{} that we use. \Cref{sec:mst-expected-opt} proves the sharp bound on the weight of a minimum spanning tree via the cut relaxation and the integrality gap (\Cref{thm:mst-sharp-bound}). \Cref{sec:o1} proves Theorem~\ref{thm:main-o1}: after reducing to the case where $n$ is a power of two, it decomposes the per-step cost into an intra-cluster term and an inter-cluster term and bounds the two terms in turn. \Cref{sec:unknown} presents \EFB{} and proves \Cref{thm:main-unknown}. Proofs of the lemmas that we import from~\cite{GGPW19} are deferred to the appendix. The appendix also contains the table containing all notations appearing in this paper.
\section{Proof Overview}
\label{sec:overview}

Before developing the formal toolkit in Section~\ref{sec:prelim}, we give a high-level overview of the proofs of Theorems~\ref{thm:main-o1} and~\ref{thm:main-unknown}. Both proofs phrase online matching decisions in terms of \emph{optimal transport}. For any non-empty finite multiset $A \subseteq X$, let $\cU_A$ denote the uniform probability distribution placing weight $1/|A|$ on each element of $A$ (with $\delta_x := \cU_{\{x\}}$ for a single point $x$). Given two probability distributions $\mu$ and $\nu$ on $X$, a \emph{transport plan} $\pi \in \Pi(\mu, \nu)$ specifies how much probability mass $\pi(x, y) \ge 0$ is moved from each $x \in X$ to each $y \in X$ to transform $\mu$ into $\nu$, and their \emph{Earth Mover's distance} $W_1(\mu, \nu) = \min_{\pi \in \Pi(\mu, \nu)} \sum_{x, y} \pi(x, y)\,d(x, y)$ is the minimum expected cost of such a plan (Section~\ref{sec:prelim}). For any $n$-element multisets $R_n$ and $S$, $n \cdot W_1(\cU_{R_n}, \cU_S)$ equals the minimum-cost perfect matching cost $\OPT(R_n, S)$.

\subsection{Question 1: \texorpdfstring{$O(1)$}{O(1)}-Competitiveness on General Metrics (Theorem~\ref{thm:main-o1})}
\label{subsec:overview-q1}

When the arrival distribution $\cD$ is known, the \FB{} algorithm matches each arriving request $r_t \sim \cD$ to the remaining $k = n - t + 1$ free servers $S_k$ by sampling a server according to a minimum-cost transport plan $\pi^{(t)} \in \Pi(\cU_{S_k}, \cD)$.

\begin{algorithm}[t]
\caption{\FB{}$(\cD)$ (for a Known Arrival Distribution $\cD$)~\cite{GGPW19}}
\label{alg:fb}
\vspace{0.3em}
\noindent\textbf{Input:} Server multiset $S \subseteq X$ of size $n$ in metric space $(X, d)$, and known arrival distribution $\cD$ over $X$.
\begin{enumerate}[leftmargin=2em,label=\arabic*.,itemsep=0.25em,topsep=0.3em]
    \item Initialize free server set $S_n \gets S$.
    \item \textbf{for} $t = 1, 2, \dots, n$ \textbf{do}
    \begin{enumerate}[leftmargin=1.8em,label=(\alph*),itemsep=0.2em,topsep=0.15em]
        \item Let $k \gets n - t + 1$.
        \item Compute a minimum-cost transport plan $\pi^{(t)} \in \Pi(\cU_{S_k}, \cD)$ achieving $W_1(\cU_{S_k}, \cD)$.
        \item Observe arriving request $r_t \sim \cD$ (so $\cD(r_t) > 0$).
        \item Sample a server $s_t \in S_k$ with probability $\Pr(s_t = s \mid S_k, r_t) = \frac{\pi^{(t)}(s, r_t)}{\cD(r_t)}$.
        \item Match $r_t$ to $s_t$, and update $S_{k-1} \gets S_k \setminus \{s_t\}$.
    \end{enumerate}
\end{enumerate}
\vspace{0.1em}
\end{algorithm}

\paragraph{Reduction to uniform server sampling.}
We follow the analysis of \FB{} in~\cite{GGPW19}. Because each point $x \in X$ arrives with probability $\cD(x)$, Step~2(d) of Algorithm~\ref{alg:fb} removes each free server $s \in S_k$ with probability $\sum_x \pi^{(t)}(s, x) = \frac{1}{k}$. Hence $S_k$ is always a uniformly random $k$-subset of $S$, and the expected step-$t$ cost is $\E[W_1(\cU_{S_k}, \cD)]$ (Lemma~\ref{lem:ggpw-structure}). By the triangle inequality through $\cU_S$ and $\E[\OPT_{\cU_S}] \le 2\,\E[\OPT_{\cD}]$ (Lemma~\ref{lem:opt-comparison}), Theorem~\ref{thm:main-o1} reduces to showing that for every $n$-point metric space $(S, d)$,
\[
    \sum_{k=1}^n u_k(S) \;\le\; O(1) \cdot \E[\OPT_{\cU_S}],
    \qquad \text{where} \qquad
    u_k(S) \;:=\; \E_{S_k}\bigl[W_1(\cU_{S_k}, \cU_S)\bigr].
\]
By the Replacement Lemma (Lemma~\ref{lem:replacement-mono}(b)), it suffices to bound the with-replacement version $\tilde{u}_k(S) := \E_{\hat{S}_k}[W_1(\cU_{\hat{S}_k}, \cU_S)] \ge u_k(S)$.

The analysis of~\cite{GGPW19} bounds $u_k(S)$ separately in two ranges of $k$. For large $k$ ($k \ge n/\log^2 n$), it works directly in $(S, d)$ and compares $S_k$ with the $k$ individual servers that receive the most requests~\cite[Lemma~3.8]{GGPW19}. For small $k$ ($k \le n/\log^2 n$), it passes to an implicit probabilistic tree embedding~\cite{FRT04} and bounds $u_k(S)$ on the tree~\cite[Lemma~3.1]{GGPW19}. Our analysis replaces both by a single argument that works directly in $(S, d)$ for every $k$, using clusters of servers instead of individual servers.

\paragraph{Core idea: Two-Stage Clustered Decomposition directly in $(S, d)$.}
We work \textbf{directly in the original metric space $(S, d)$} for all $k$ (incurring zero tree-embedding distortion) and group servers into $B = n/m$ equal-sized \textbf{clusters $C_1, \dots, C_B \subseteq S$ of sublinear size $m \ge 2$}. Whereas a single server's request load follows a $\mathrm{Poisson}(1)$ tail, a cluster of $m$ servers receives $\mathrm{Bin}(n, m/n)$ requests, which fluctuates around its capacity $m$ on the scale $\Theta(\sqrt{m})$. 
In the Two-Stage Clustered Decomposition (\Cref{lem:two-stage-decomp}), we split the cost of transporting a with-replacement sample $\hat{S}_k = (\hat{s}_1, \dots, \hat{s}_k)$ to $\cU_S$ into two stages:
\begin{itemize}[leftmargin=2em,itemsep=0.15em]
    \item \textbf{Stage~1 (Intra-cluster smoothing):} Spread each sampled server's point mass uniformly across its own cluster of size $m$.
    \item \textbf{Stage~2 (Inter-cluster transport):} Transport the resulting $k$-cluster mixture to the uniform distribution $\cU_S$ over all $B$ clusters.
\end{itemize}

\paragraph{Warm-up: Divisible cluster size and distinct cluster draws.}
To see how Stage~1 and Stage~2 interact---and why they dictate a sublinear cluster size $m \asymp (n/k)^{2/3}$---suppose first that $n$ is divisible by $m$ (so $S$ partitions into $B = n/m$ clusters of exact size $m$) and that the $k$ sampled servers fall in $k$ \emph{distinct} clusters $A \subseteq [B]$ ($|A| = k$, with no collisions):
\begin{itemize}[leftmargin=2em,itemsep=0.2em]
    \item \emph{Stage~1 via MST tour slicing (\Cref{alg:cluster-partition}):} 
    We construct a Minimum Spanning Tree $T = \MST(S, d)$, order $S$ \old{along a depth-first tour of $T$ (traversing every edge twice, for total tour length at most $2\,\wt(T)$)} \new{in the order in which a depth-first traversal of $T$ first visits them (the pre-order; consecutive servers are at total distance at most $2\,\wt(T)$, since the traversal walks every edge of $T$ twice)}, and slice the tour into $B = n/m$ contiguous blocks $C_1, \dots, C_B$ of size $m$. Because these $B$ cluster segments are disjoint portions of the tour, their average length is at most $\frac{2}{B}\,\wt(T) = \frac{2m}{n}\,\wt(T)$. Once we show that $\wt(T) \le \old{O(1)}\new{e}\,\E[\OPT_{\cU_S}]$ (\Cref{thm:mst-sharp-bound}), smoothing a uniformly random server across its own cluster costs at most $O\bigl(\frac{m}{n}\bigr)\,\E[\OPT_{\cU_S}]$ (Lemma~\ref{lem:stage1-bound}).
    \item \emph{Stage~2 via cluster load fluctuations (Lemmas~\ref{lem:clustered-bb} and~\ref{lem:top-r-load}):} Under $R_n \sim \cU_S^n$, each cluster load $X_a \sim \mathrm{Bin}(n, m/n)$ fluctuates around $m$ by $\Theta(\sqrt{m})$, so for $k\le B/2$, the $k$ most loaded clusters $A^* \subseteq [B]$ collectively receive an expected excess of $\Delta_k = \Omega(k\sqrt{m})$ requests above their capacity $km$. Any offline matching $\OPT(R_n, S)$ must ship at least these $L_k-km$ excess requests---whose expectation is $\Delta_k$---out of $\bigcup_{a \in A^*} C_a$ to the remaining clusters. Moreover, by symmetry across the $B$ equal-sized clusters, $A^*$ is a uniformly random $k$-subset of $[B]$, and averaging the empirical request distribution $\cU_{R_n}$ conditioned on $\{A^* = A\}$ cancels intra-cluster noise to leave a net excess of $\frac{\Delta_k}{n}$ on the $k$ clusters in $A$. Consequently, the expected cost of transporting the $k$ clusters in $A$ to $\cU_S$ is at most $\frac{1}{\Delta_k}\,\E[\OPT_{\cU_S}] = O\bigl(\frac{1}{k\sqrt{m}}\bigr)\,\E[\OPT_{\cU_S}]$.
    \item \emph{Balancing Stage~1 and Stage~2:} Notice the trade-off in the cluster size $m$: larger clusters strengthen the Stage~2 balls-and-bins bound $O\bigl(\frac{1}{k\sqrt{m}}\bigr)$, whereas smaller clusters shrink the Stage~1 smoothing bound $O\bigl(\frac{m}{n}\bigr)$. Equating $\frac{m}{n} \asymp \frac{1}{k\sqrt{m}}$ selects a cluster size $m \asymp (n/k)^{2/3}$ (so $B = n/m \asymp n^{1/3}k^{2/3} \gg k$), under which both stages cost $O\bigl(\frac{1}{n^{1/3}k^{2/3}}\bigr)\,\E[\OPT_{\cU_S}]$. Because the exponent $2/3$ is strictly smaller than $1$, summing over $k = 1, \dots, n$ converges to $\sum_{k=1}^n \frac{1}{n^{1/3}k^{2/3}} = O(1)$.
\end{itemize}

\paragraph{From the warm-up to the full proof.}
Turning this warm-up into a complete proof for every finite metric space $(S, d)$ requires three additional ingredients, developed in \Cref{sec:mst-expected-opt} and Sections~\ref{subsec:pow2}--\ref{subsec:stage2}:
\begin{enumerate}[leftmargin=2.2em,label=\textbf{(\arabic*)},itemsep=0.35em]
    \item \textbf{Bounding the MST weight via the cut relaxation LP (\Cref{thm:mst-sharp-bound}, \Cref{sec:mst-expected-opt}):}
    To prove $\wt(T) \le e\,\E[\OPT_{\cU_S}]$, we compare both sides through the undirected cut relaxation of the minimum spanning tree problem. Every perfect matching between $R_n$ and $S$ crosses each cut $(U, S\setminus U)$ at least $\bigl|X(U) - |U|\bigr|$ times, where $X(U) \sim \mathrm{Bin}(n, |U|/n)$ is the number of requests landing in $U$. This expected crossing number is at least $b := 2(1-\frac1n)^n$. Hence the expected optimal matching, scaled down by $b$, is a feasible fractional solution of value $\E[\OPT_{\cU_S}]/b$. The integrality gap of at most $2(1-\frac1n)$ for this relaxation, due to \cite{GW95,AKR95}, then gives
    \(
        \wt(T) \le \frac{2(1-1/n)}{b}\,\E[\OPT_{\cU_S}] \le e\,\E[\OPT_{\cU_S}].
    \)
    This constant is tight due to the lower bound from the $n$-point uniform metric. The bound may be of independent interest.

    \item \textbf{Handling cluster collisions via multiplicity peeling (Lemmas~\ref{lem:stage2-peeling} and~\ref{lem:per-step-uk}, \S\ref{subsec:two-stage} and~\S\ref{subsec:stage2}):}
    When $k$ cluster indices $\hat{A}_k = (a_1, \dots, a_k)$ are drawn with replacement from $[B]$ (with $m \asymp (n/k)^{2/3}$ chosen as a power of~$2$ dividing $n = 2^\ell$ so that $B = n/m \gg k$), some clusters may be drawn more than once. We slice $\hat{A}_k$ into \emph{multiplicity layers} $A^{(j)}$ (the set of clusters drawn at least $j$ times, of random size $K_j = |A^{(j)}|$); conditioned on $K_j = r \ge 1$, symmetry ensures that $A^{(j)}$ is a uniformly random $r$-subset of clusters.
    \begin{itemize}[leftmargin=1.8em,itemsep=0.15em]
        \item For the first layer $j = 1$ (the $K_1$ distinct clusters hit by $\hat{A}_k$), applying Lemmas~\ref{lem:clustered-bb} and~\ref{lem:top-r-load} ($\Delta_r = \Omega(r\sqrt{m})$) given $K_1 = r$ bounds the transport cost by $O\bigl(1/(r\sqrt{m})\bigr)\,\E[\OPT_{\cU_S}]$; multiplying by the layer's \old{probability weight} \new{mixture weight} $r/k$ cancels $r$ and gives $O\bigl(1/(k\sqrt{m})\bigr)\,\E[\OPT_{\cU_S}] = O\bigl(n^{-1/3}k^{-2/3}\bigr)\,\E[\OPT_{\cU_S}]$.
        \item Applying that same bound across all layers $j \ge 1$ would pay a factor of $\sum_{j \ge 1} \Pr(K_j \ge 1)$---the expected maximum bin load---and reintroduce a logarithmic loss. Instead, for every collision layer $j \ge 2$, each cluster in $A^{(j)}$ is drawn at least twice ($K_j \le \lfloor k/2 \rfloor < k$), and the Replacement Lemma together with Jensen's inequality bounds the cost of transporting $K_j$ uniformly random clusters by $\tilde{u}_{K_j}(S)$. This gives recurrence relation \eqref{eq:uk-step}:
        \begin{equation*}
            \tilde{u}_k(S)
            \le
            O({n^{-1/3}k^{-2/3}})\,\E[\OPT_{\cU_S}]
            +\sum_{j=2}^k
            \E\left[\frac{K_j}{k}\tilde{u}_{K_j}(S)\right].
        \end{equation*}
        Then our choice of parameters ensures $\tilde{u}_k(S)\le O({n^{-1/3}k^{-2/3}})\,\E[\OPT_{\cU_S}]$.
    \end{itemize}

    \item \textbf{Power-of-two server reduction (Lemma~\ref{lem:pow2-red}, Section~\ref{subsec:pow2}):}
    Finally, when $n$ is not a power of two, let $n' := 2^{\lfloor \log_2 n \rfloor} \in (n/2, n]$. Routing each $k$-subset $S_k \subseteq S$ ($k \le n'$) through a uniformly random $n'$-subset $S' \subseteq S$ via the triangle inequality (and using subset monotonicity for $k > n'$) reduces any $n$-server instance to the power-of-two sub-instance $S'$, while subsampling $n'$ of the $n$ requests in $R_n \sim \cU_S^n$ bounds both the bridge cost $n\,u_{n'}(S)$ and the expected sub-instance optimum $\E_{S'}\bigl[\E[\OPT_{\cU_{S'}}(S')]\bigr]$ by $O(1)\,\E[\OPT_{\cU_S}]$.
\end{enumerate}

\subsection{Question 2: Unknown Arrival Distributions (Theorem~\ref{thm:main-unknown})}
\label{subsec:overview-q2}

\paragraph{Why naive empirical plug-in fails.}
When the arrival distribution $\cD$ is unknown upfront, the algorithm cannot compute a transport plan to $\cD$. Ignoring past requests altogether and matching each arriving request $r_t$ to a uniformly random free server in $S_k$ fails even on a $2$-point metric space with $n/2$ servers at each point and uniform arrivals (paying $\Theta(n)$ in expectation while $\E[\OPT_{\cD}] = \Theta(\sqrt{n})$). On the other hand, if we try to estimate $\cD$ at step $t$ from the \emph{strictly past} requests $(r_1, \dots, r_{t-1})$ (or from an initial ``burn-in'' prefix of arrivals), two problems arise: the new request $r_t$ may land at a point outside $\{r_1, \dots, r_{t-1}\}$, \old{and worse, past matching decisions correlate the remaining free server set $S_k$ with $(r_1, \dots, r_{t-1})$, destroying the uniform-randomness invariant $S_k \sim \cU_k(S)$ that underpins \FB{}~\cite{GGPW19}.} \new{and, more importantly, the estimate no longer cancels the denominator of the sampling rule, so past matching decisions can correlate the remaining free server set $S_k$ with $(r_1, \dots, r_{t-1})$, and the uniform-randomness invariant $S_k \sim \cU_k(S)$ that underpins \FB{}~\cite{GGPW19} need not hold.}

\paragraph{Key idea: Inclusive empirical distributions and i.i.d.\ symmetry (\Cref{subsec:efb}).}
As shown in \Cref{subsec:overview-q1}, when $\cD$ is known, \FB{} keeps $S_k$ uniformly random because the arrival probability $\cD(x)$ cancels the denominator in $\pi^{(t)}(s, x)/\cD(x)$. When $\cD$ is unknown, the \EFB{} algorithm (Algorithm~\ref{alg:efb}) restores this exact cancellation without any burn-in phase: at step $t$, \emph{after} observing the arriving request $r_t$, it forms the \textbf{inclusive empirical distribution}
\[
    \hat{\cD}_t \;:=\; \frac{1}{t}\sum_{i=1}^t \delta_{r_i}
\]
of all $t$ requests seen so far (including $r_t$ itself), computes a minimum-cost transport plan $\pi^{(t)}$ between $\cU_{S_k}$ and $\hat{\cD}_t$ using a fixed deterministic tie-breaking rule that depends \emph{only} on $S_k$ and the unordered multiset $\{r_1, \dots, r_t\}$, and matches $r_t$ to server $s \in S_k$ with probability $\pi^{(t)}(s, r_t)/\hat{\cD}_t(r_t)$.

Why does including $r_t$ in $\hat{\cD}_t$ keep $S_k$ uniformly random and independent of the empirical distribution? Because $r_1, \dots, r_t$ are drawn i.i.d.\ from $\cD$, their joint distribution is symmetric across arrival orders: conditioned on the \emph{unordered multiset} $\{r_1, \dots, r_t\}$ (equivalently, $\hat{\cD}_t$), each of the $t$ elements in $\{r_1, \dots, r_t\}$ is equally likely to be the one that arrived at step $t$, so $\Pr(r_t = x \mid \hat{\cD}_t) = \hat{\cD}_t(x)$. Therefore, if $S_k$ is independent of $\hat{\cD}_{t-1}$ (and hence independent of $(\hat{\cD}_t, r_t)$), the conditional probability that any free server $s \in S_k$ is matched at step $t$ is
\[
    \sum_{x : \hat{\cD}_t(x) > 0} \underbrace{\Pr\bigl(r_t = x \;\bigm|\; S_k,\, \hat{\cD}_t\bigr)}_{=\;\hat{\cD}_t(x)} \cdot \frac{\pi^{(t)}(s, x)}{\hat{\cD}_t(x)} \;=\; \sum_{x \in X} \pi^{(t)}(s, x) \;=\; \frac{1}{k}.
\]
Notice that $\hat{\cD}_t(x)$ \emph{cancels out completely}: every server in $S_k$ is matched with probability $1/k$, regardless of $\hat{\cD}_t$. By induction on $t$, the free server set $S_k$ remains a uniformly random $k$-subset of $S$ independent of $\hat{\cD}_t$ at every step, and the expected step-$t$ matching cost is $\E[W_1(\cU_{S_k}, \hat{\cD}_t)]$ (Lemma~\ref{lem:exchangeable-structure}).

\paragraph{Three-hop time-reversal coupling (\Cref{subsec:time-reversal}).}
We view the first $t$ requests backward in time as a uniformly random $t$-element subset $\tilde{R}_t$ of the full $n$-request sequence $R_n \sim \cD^n$. Under any offline optimal matching $\sigma^* : [n] \to [n]$ between $R_n$ and $S$, the $t$ servers $\tilde{S}_t \subseteq S$ matched to $\tilde{R}_t$ form a uniformly random $t$-subset of $S$. Applying the triangle inequality along the three-hop chain $\cU_{S_k} \to \cU_S \to \cU_{\tilde{S}_t} \to \cU_{\tilde{R}_t}$ bounds the step-$t$ cost solely in terms of uniform server subsets and $\OPT_{\cD}$:
\[
    \E\bigl[W_1(\cU_{S_k}, \hat{\cD}_t)\bigr] \;\le\; \underbrace{\E\bigl[W_1(\cU_{S_k}, \cU_S)\bigr]}_{=\;u_k(S)} \;+\; \underbrace{\E\bigl[W_1(\cU_S, \cU_{\tilde{S}_t})\bigr]}_{=\;u_t(S)} \;+\; \underbrace{\E\bigl[W_1(\cU_{\tilde{S}_t}, \cU_{\tilde{R}_t})\bigr]}_{\le\;\frac{1}{n}\,\E[\OPT_{\cD}]}.
\]
\new{We subsample an independent random index set rather than taking the first $t$ requests themselves, because $\sigma^*$ depends on the whole sequence $R_n$: conditioned on $R_n$, the servers that $\sigma^*$ matches to the fixed indices $1, \dots, t$ are deterministic, whereas the servers matched to an independent uniformly random $t$-subset of indices form a uniformly random $t$-subset of $S$ by construction, without requiring $\sigma^*$ to be chosen symmetrically with respect to the arrival order.}
Summing over $t = 1, \dots, n$ (while $k = n - t + 1$ runs from $n$ down to $1$) yields Theorem~\ref{thm:main-unknown}: $\E[\EFB(\cD)] \le 2\sum_{k=1}^n u_k(S) + \E[\OPT_{\cD}]$. Combining this reduction with Theorem~\ref{thm:main-o1} immediately proves $O(1)$-competitiveness for unknown distributions (Corollary~\ref{cor:unknown-o1}).

\section{Preliminaries}
\label{sec:prelim}

In this section, we collect our standing notation, basic definitions from optimal transport, and foundational lemmas on sampling and the \FB{} algorithm.

\paragraph{Standing Notation and Conventions.}
For any integer $m \ge 1$, we write $[m] := \{1, \dots, m\}$. Throughout, $(X, d)$ is a finite metric space, and $S = \{s_1, \dots, s_n\} \subseteq X$ is a multiset of $n \ge 1$ servers in which co-located points in $X$ are distinguished by their indices in $[n]$ (so subsets of $S$ are subsets of these $n$ indexed servers). For any event or condition $E$, we write $\mathbf{1}_{\{E\}}$ for its indicator, equal to $1$ if $E$ holds and $0$ otherwise. For an integer $n \ge 1$ and $p \in [0, 1]$, we write $Z \sim \mathrm{Bin}(n, p)$ for a binomial random variable (the sum of $n$ independent $\{0,1\}$-valued trials each succeeding with probability $p$), which has mean $\E[Z] = np$ and variance $\mathrm{Var}(Z) = np(1-p)$.

\subsection{Probability Distributions and Earth Mover's Distance \texorpdfstring{($W_1$)}{(W1)}}

A \emph{probability distribution} $\mu$ on $X$ assigns a non-negative weight $\mu(x) \ge 0$ to each point $x \in X$ such that $\sum_{x \in X} \mu(x) = 1$.

\begin{definition}[Uniform Distributions and Point Masses]
For any non-empty finite multiset $A \subseteq X$, the \emph{uniform distribution} on $A$, denoted $\cU_A$, places equal weight $1/|A|$ on each element of $A$:
\[
    \cU_A(x) \;:=\; \frac{|\{a \in A : a = x\}|}{|A|} \qquad \text{for all } x \in X.
\]
For a single point $x \in X$, we write $\delta_x := \cU_{\{x\}}$ for the \emph{point mass} (or Dirac delta) placing weight $1$ at $x$ and $0$ everywhere else. Thus $\cU_A = \frac{1}{|A|}\sum_{a \in A} \delta_a$. For a finite sequence $A$ of points of $X$, $\cU_A$ denotes the uniform distribution on the multiset of its entries.
\end{definition}

\begin{definition}[Transport Plans and Earth Mover's Distance $W_1$]
Given two probability distributions $\mu$ and $\nu$ on $X$, a \emph{transport plan} (also called a \emph{coupling}) from $\mu$ to $\nu$ is a function $\pi : X \times X \to \R_{\ge 0}$ specifying how much probability mass $\pi(x, y)$ is moved from each point $x$ to each point $y$, such that
\[
    \sum_{y \in X} \pi(x, y) \;=\; \mu(x) \quad \text{for all } x \in X \qquad \text{and} \qquad
    \sum_{x \in X} \pi(x, y) \;=\; \nu(y) \quad \text{for all } y \in X.
\]
We denote the set of all transport plans from $\mu$ to $\nu$ by $\Pi(\mu, \nu)$. The \emph{Earth Mover's distance} (or \emph{$1$-Wasserstein distance}) between $\mu$ and $\nu$ is the minimum cost required to move $\mu$ into $\nu$:
\[
    W_1(\mu, \nu) \;:=\; \min_{\pi \in \Pi(\mu, \nu)} \sum_{x, y \in X} \pi(x, y)\, d(x, y).
\]
\end{definition}
\new{When a marginal is uniform on a multiset, we may retain the labels of its elements and compute transport costs using their locations in $X$. In particular, a transport plan with marginal $\cU_A$ on an indexed server multiset $A$ has one row per server, each of total mass $1/|A|$; in this representation, $\cU_A(s)=1/|A|$ for each indexed server $s\in A$. Aggregating co-located labels recovers the corresponding distribution on $X$ and does not change the minimum transport cost. When requests are sampled from $\cU_S$, we likewise retain their independently sampled uniform server labels, and membership in server subsets, cuts, or clusters refers to these labels. Thus we may work on the indexed servers with distance inherited from $X$, allowing distinct labels to have distance zero.}

We will repeatedly use four standard properties of $W_1(\mu, \nu)$:
\begin{enumerate}[leftmargin=2em,label=\textbf{(P\arabic*)}]
    \item \label{prop:w1-triangle} \textbf{Triangle Inequality}~\cite[Chapter~6]{Vil09}\textbf{:} For any three distributions $\mu, \nu, \lambda$ on $X$, we have $W_1(\mu, \lambda) \le W_1(\mu, \nu) + W_1(\nu, \lambda)$.

    \item \label{prop:w1-convex} \textbf{Convexity and Jensen's Inequality}~\cite[Theorem~4.8]{Vil09}\textbf{:} For any weights $\alpha_1, \dots, \alpha_m \ge 0$ summing to $1$ and distributions $\mu_i, \nu_i$ on $X$, we have $W_1\bigl(\sum_i \alpha_i \mu_i, \sum_i \alpha_i \nu_i\bigr) \le \sum_i \alpha_i W_1(\mu_i, \nu_i)$. In particular, for any random distribution $\hat{\mu}$ and fixed distribution $\nu$, Jensen's inequality gives
    \[
        W_1\bigl(\E[\hat{\mu}],\, \nu\bigr) \;\le\; \E\bigl[W_1(\hat{\mu},\, \nu)\bigr].
    \]

    \item \label{prop:w1-matching} \textbf{Equivalence to Minimum-Cost Perfect Matching}~\cite[Proposition~2.1]{PC19}\textbf{:} By the Birkhoff--von Neumann integrality theorem for bipartite matching, for any $n$-request sequence $R_n = (r_1, \dots, r_n)$ and $n$-server multiset $S = \{s_1, \dots, s_n\}$, we have
    \[
        \OPT(R_n, S) \;=\; n \cdot W_1(\cU_{R_n}, \cU_S).
    \]

    \item \label{prop:w1-shortcut} \textbf{The Shortcut Property}~\cite[Lemma~3.5]{GGPW19} \textbf{(Moving Only Net Excess to Net Deficit):} Suppose $\mu \ne \nu$. At each point $x \in X$, let $\mu^+(x) := \max(\mu(x) - \nu(x), 0)$ be the \emph{net excess} at $x$, and $\mu^-(x) := \max(\nu(x) - \mu(x), 0)$ be the \emph{net deficit} at $x$. Let $\alpha := \sum_{x \in X} \mu^+(x) = \sum_{x \in X} \mu^-(x) > 0$ be the total excess weight. By~\cite[Lemma~3.5]{GGPW19}, there exists an optimal transport plan $\pi \in \Pi(\mu, \nu)$ keeping $\pi(x, x) = \min(\mu(x), \nu(x))$ units of shared mass stationary at every $x \in X$ and transporting only $\mu^+$ to $\mu^-$, so
    \[
        W_1(\mu, \nu) \;=\; \alpha \cdot W_1\bigl(\mu^+/\alpha,\; \mu^-/\alpha\bigr).
    \]
\end{enumerate}

\subsection{Sampling With and Without Replacement, and the \texorpdfstring{\FB{}}{Fair-Bias} Algorithm}

For any non-empty finite set or multiset $A$ and integer $k \ge 1$, we use two ways of sampling $k$ elements from $A$:
\begin{itemize}[leftmargin=2em]
    \item \textbf{Without replacement ($A_k \sim \cU_k(A)$, for $1 \le k \le |A|$):} the subset $A_k$ is a uniformly random $k$-element subset of $A$ (each of the $\binom{|A|}{k}$ subsets of size $k$ is chosen with probability $\binom{|A|}{k}^{-1}$). We also let $\cU_0(A)$ denote the distribution concentrated on the empty subset of $A$.
    \item \textbf{With replacement ($\hat{A}_k \sim \cI_k(A)$):} the sample $\hat{A}_k = (a_1, \dots, a_k)$ is a sequence (or multiset) of $k$ elements drawn independently and uniformly at random from $A$ (each $a_i$ is chosen from $A$ with probability $1/|A|$, so the same element may be drawn multiple times).
\end{itemize}
Specializing to our $n$-server multiset $S$ ($1 \le k \le n$), we denote the expected Earth Mover's distance from $\cU_{S_k}$ (or $\cU_{\hat{S}_k}$) to $\cU_S$ by
\begin{equation}
\label{eq:uk-def}
    u_k(S) \;:=\; \E_{S_k \sim \cU_k(S)}\bigl[W_1(\cU_{S_k}, \cU_S)\bigr], \qquad
    \tilde{u}_k(S) \;:=\; \E_{\hat{S}_k \sim \cI_k(S)}\bigl[W_1(\cU_{\hat{S}_k}, \cU_S)\bigr].
\end{equation}
We also set $\tilde{u}_0(S) := 0$.

We now record three structural lemmas from~\cite{GGPW19} for the \FB{}$(\cD)$ algorithm (Algorithm~\ref{alg:fb}) and uniform server sampling. Since our statements slightly generalize those in~\cite{GGPW19} (which stated Lemma~\ref{lem:ggpw-structure}(a) for $\cD = \cU_S$ and handled general $\cD$ via a transshipment wrapper), we include short self-contained proofs in Appendix~\ref{app:prelim-proofs}.

\begin{lemma}[Structure of \FB{} {\cite[Lemmas~2.1 and~2.2]{GGPW19}}]
\label{lem:ggpw-structure}
For every server multiset $S \subseteq X$ of size $n \ge 1$ and every arrival distribution $\cD$ over $X$, under \FB{}$(\cD)$ (Algorithm~\ref{alg:fb}), the following hold:
\begin{enumerate}[leftmargin=2em,label=(\alph*)]
    \item At every step $t \in [n]$ (with $k = n - t + 1$), the remaining free server set $S_k$ is a uniformly random $k$-subset of $S$ ($S_k \sim \cU_k(S)$), and the expected matching cost at step $t$ is $\E_{S_k \sim \cU_k(S)}[W_1(\cU_{S_k}, \cD)]$.
    \item In particular, $\E[\FB(\cU_S)] = \sum_{k=1}^n u_k(S)$, and for any distribution $\cD$ over $X$,
    \[
        \E[\FB(\cD)] \;\le\; \sum_{k=1}^n u_k(S) + \E[\OPT_{\cD}].
    \]
\end{enumerate}
\end{lemma}

\begin{lemma}[Monotonicity and Replacement {\cite[Lemmas~4.3 and~B.1]{GGPW19}}]
\label{lem:replacement-mono}
For every server multiset $S \subseteq X$ of size $n \ge 1$:
\begin{enumerate}[leftmargin=2em,label=(\alph*)]
    \item\label{prop:mono} \emph{(Monotonicity)} Larger subsets are closer to $S$: for $1 \le k \le k' \le n$, we have $u_{k'}(S) \le u_k(S)$.
    \item\label{prop:replace} \emph{(Replacement Lemma)} Sampling with replacement costs at least as much as sampling without replacement: for all $1 \le k \le n$, we have $u_k(S) \le \tilde{u}_k(S)$. More generally, for any distributions $\mu_1, \dots, \mu_N$ and $\nu$ on $X$ and any $1 \le k \le N$,
    \[
        \E_{A \sim \cU_k([N])}\!\left[W_1\!\left(\frac{1}{k}\sum_{a \in A} \mu_a,\, \nu\right)\right] \;\le\; \E_{(a_1, \dots, a_k) \sim \cI_k([N])}\!\left[W_1\!\left(\frac{1}{k}\sum_{i=1}^k \mu_{a_i},\, \nu\right)\right].
    \]
\end{enumerate}
\end{lemma}

\begin{lemma}[Comparison of Offline Optima {\cite[Appendix~A, Equations~(4)--(5)]{GGPW19}}]
\label{lem:opt-comparison}
For every server multiset $S \subseteq X$ of size $n \ge 1$ and every arrival distribution $\cD$ over $X$, we have $\E[\OPT_{\cU_S}] \le 2\,\E[\OPT_{\cD}]$.
\end{lemma}

Finally, to lower-bound the random fluctuations of cluster loads in Lemma~\ref{lem:top-r-load}, we use a lower bound on the mean absolute deviation of a binomial random variable due to Berend and Kontorovich~\cite[Theorem~1]{BK13} (noting that $1 - p \ge 1/2$ for $p \le 1/2$).

\begin{lemma}[Binomial Mean Absolute Deviation {\cite[Theorem~1]{BK13}}]
\label{lem:bin-mad}
Let $Z \sim \mathrm{Bin}(n, p)$ with $n \ge 2$ and $\frac{1}{n} \le p \le \frac{1}{2}$. Then
\[
    \E\bigl[|Z - np|\bigr] \;\ge\; \sqrt{\frac{np(1-p)}{2}} \;\ge\; \frac{1}{2}\sqrt{np}.
\]
\end{lemma}

\section{Minimum Spanning Tree and the Expected Offline Optimum}
\label{sec:mst-expected-opt}

Let $n:=|S|\ge 2$, $E:=\binom{S}{2}$, and
$w_{\{u,v\}}:=d(u,v)$ for $\{u,v\}\in E$.
Write
$x(F)=\sum_{e\in F}x_e$ for $F\subseteq E$.
Let $T:=\MST(S,d)$ be a minimum spanning tree of $(S,d)$, with weight
$\wt(T):=\sum_{e\in T}w_e$.
The main result of this section is the following bound.
\begin{theorem}
\label{thm:mst-sharp-bound}
The minimum spanning tree $T$ satisfies
\[
    \wt(T)
    \le C_n\,\mathbb{E}[\mathrm{OPT}_{\mathcal U_S}]
    \qquad \text{where} \qquad
    C_n:=\left(1+\frac{1}{n-1}\right)^{n-1}.
\]
\end{theorem}
Since $C_n<e$, this implies
$\wt(T)\le e\,\mathbb{E}[\mathrm{OPT}_{\mathcal U_S}]$, which also holds trivially for $n=1$.
This is the bound on the minimum spanning tree used in the Stage~1 analysis
(\Cref{lem:stage1-bound}).
For each fixed $n$, the bound is attained with equality by the $n$-point uniform metric
(\Cref{lem:mst-uniform-lower}), so $C_n$ cannot be replaced by a smaller constant valid
for all $n$-point metrics.

\paragraph{The cut relaxation.}
For $U\subseteq S$, let $\delta(U)$ denote the set of
edges with exactly one endpoint in $U$. Also, let $\mathcal C:=\{U:\varnothing\ne U\subsetneq S\}$.
The undirected cut relaxation of the minimum spanning tree problem and its dual are:
\par\noindent
\begin{minipage}[t]{0.48\linewidth}
\begin{equation}
\begin{gathered}
    \mathrm{PRIMAL}\\
    \begin{aligned}
    \min\quad &\sum_{e\in E}w_ex_e\\
    \text{s.t.}\quad
        &x(\delta(U))\ge 1
        &&\forall U\in\mathcal C,\\
        &x_e\ge 0
        &&\forall e\in E.
    \end{aligned}
\end{gathered}
\label{eq:mst-cut-lp}
\end{equation}
\end{minipage}\hfill
\begin{minipage}[t]{0.48\linewidth}
\begin{equation}
\begin{gathered}
    \mathrm{DUAL}\\
    \begin{aligned}
    \max\quad &\sum_{U\in\mathcal C}y_U\\
    \text{s.t.}\quad
        &\sum_{\substack{U\in\mathcal C\\e\in\delta(U)}}y_U
          \le w_e
        &&\forall e\in E,\\
        &y_U\ge 0
        &&\forall U\in\mathcal C.
    \end{aligned}
\end{gathered}
\label{eq:mst-cut-dual}
\end{equation}
\end{minipage}
\par\medskip
Denote their common optimal value by $\mathrm{LP}$.
\begin{theorem}[Theorem 2.4~\cite{GW95} (see also \cite{AKR95})]
\label{thm:mst-gw-gap}
The integrality gap of \eqref{eq:mst-cut-lp} is at most $2(1-1/n)$. In particular,
\begin{equation*}
    \wt(T)\le 2\left(1-\frac1n\right)\mathrm{LP}.
\end{equation*}
\end{theorem}
Their primal--dual algorithm, run with every vertex as a terminal, returns a spanning
tree of weight at most $2(1-1/n)\,\mathrm{LP}$; the bound follows since $T$ is a minimum
spanning tree.

\paragraph{A scaled fractional solution.}
We now show that the expected optimal matching, scaled down by $b$, is feasible
for \eqref{eq:mst-cut-lp}.

\begin{lemma}
\label{lem:mst-binomial-imbalance}
For every integer $1\le k\le n-1$,
\[
    b_k:=
    \mathbb{E}\left|
        \operatorname{Bin}\left(n,\frac{k}{n}\right)-k
    \right|
    \ge b:=2\left(1-\frac1n\right)^n,
\]
with equality for $k=1$ and $k=n-1$.
\end{lemma}

\begin{proof}
By De~Moivre's formula for the mean absolute deviation of a binomial random
variable (see \cite{DZ91}), whenever $k=np$ is an integer,
$\mathbb{E}|\operatorname{Bin}(n,p)-np|=2(k+1)\binom{n}{k+1}p^{k+1}(1-p)^{n-k}$.
Substituting $p=k/n$ and using $(k+1)\binom{n}{k+1}=(n-k)\binom nk$ gives
\[
    b_k
    =\frac{2k(n-k)}{n}\binom nk
      \left(\frac{k}{n}\right)^k
      \left(\frac{n-k}{n}\right)^{n-k}.
\]
This is symmetric under $k\mapsto n-k$, and $b_1=b$. For $1\le k\le n-2$,
\[
    \frac{b_{k+1}}{b_k}=\frac{g(k)}{g(n-k-1)},
    \qquad g(t):=\left(1+\frac1t\right)^{t+1}.
\]
Since $\frac{d}{dt}\log g(t)=\log(1+1/t)-1/t<0$, $g$ is decreasing on $(0,\infty)$, so
$b_{k+1}\ge b_k$ whenever $k\le(n-1)/2$. With symmetry, the minimum is attained at
$k\in\{1,n-1\}$.
\end{proof}

\begin{lemma}
\label{lem:mst-opt-feasible}
There is a vector $x\ge 0$ with $\sum_{e\in E}w_ex_e=\mathbb{E}[\mathrm{OPT}_{\mathcal U_S}]$
and $x(\delta(U))\ge b$ for every $\varnothing\ne U\subsetneq S$. In particular, $x/b$ is
feasible for \eqref{eq:mst-cut-lp}, and
\[
    \mathrm{LP}\le\frac{1}{b}\,\mathbb{E}[\mathrm{OPT}_{\mathcal U_S}].
\]
\end{lemma}

\begin{proof}
For each realization $R\sim\mathcal U_S^n$, fix an optimal matching between $R$ and $S$.
For $e=\{u,v\}\in E$, let $x_e$ be the expected number of matched pairs whose endpoints
form $e$; pairs matched at the same server have zero cost and are omitted. By linearity
of expectation, $\sum_{e}w_ex_e=\mathbb{E}[\mathrm{OPT}_{\mathcal U_S}]$.

Fix $\varnothing\ne U\subsetneq S$ and let $X_R(U)$ be the number of requests in $U$.
Every matching has at least $|X_R(U)-|U||$ pairs crossing $\delta(U)$. Since
$X_R(U)\sim\operatorname{Bin}(n,|U|/n)$, \Cref{lem:mst-binomial-imbalance} gives
$x(\delta(U))\ge b_{|U|}\ge b$.
\end{proof}

\begin{proof}[Proof of \Cref{thm:mst-sharp-bound}]
Combining \Cref{thm:mst-gw-gap} with \Cref{lem:mst-opt-feasible},
\[
    \wt(T)
    \le \frac{2(1-1/n)}{2(1-1/n)^n}\,\mathbb{E}[\mathrm{OPT}_{\mathcal U_S}]
    =\left(1+\frac{1}{n-1}\right)^{n-1}\mathbb{E}[\mathrm{OPT}_{\mathcal U_S}]
    =C_n\,\mathbb{E}[\mathrm{OPT}_{\mathcal U_S}]. \qedhere
\]
\end{proof}

\begin{lemma}[Tightness]
\label{lem:mst-uniform-lower}
For every $n\ge 2$, the $n$-point uniform metric satisfies
\[
    \wt(T)
    =C_n\mathbb{E}[\mathrm{OPT}_{\mathcal U_S}].
\]
Consequently, $e$ is the smallest universal constant
$C$ for which
\(
    \wt(T)
    \le C\mathbb{E}[\mathrm{OPT}_{\mathcal U_S}]
\)
holds for all $n$ and all finite metrics.
\end{lemma}

\begin{proof}
In the uniform metric, normalize the common nonzero
distance to $1$, so every spanning tree has weight
$n-1$. For a request realization, let $Z$ be the
number of servers receiving no request. The total excess equals the total deficit $Z$.
Every matching must incur cost at least $Z$ to cover
this deficit, and this bound is attained by matching
locally whenever possible and assigning the excess
requests to the remaining servers. Thus $\mathrm{OPT}=Z$.

Each server receives no request with probability
$(1-1/n)^n$.
Therefore, by linearity of expectation,
\[
    \frac{\wt(T)}
         {\mathbb{E}[\mathrm{OPT}_{\mathcal U_S}]}
    =\frac{n-1}{n(1-1/n)^n}
    =\left(\frac{n}{n-1}\right)^{n-1}
    =C_n.
\]
This proves sharpness for every $n$. The optimality
of the universal constant $e$ follows from
$C_n\to e$ and \Cref{thm:mst-sharp-bound}.
\end{proof}
\section{\texorpdfstring{$O(1)$}{O(1)}-Competitiveness on General Metrics (Question 1)}
\label{sec:o1}

In this section, we prove Theorem~\ref{thm:main-o1}. By Lemmas~\ref{lem:ggpw-structure}(b) and~\ref{lem:opt-comparison}, it suffices to prove that for any $n$-point metric space $(S, d)$ (where co-located servers at distance $0$ are distinguished by their indices in $[n]$),
\[
    \sum_{k=1}^n u_k(S) \;\le\; O(1) \cdot \E[\OPT_{\cU_S}].
\]
\old{Note that~\cite[Theorem~1.1]{GGPW19} already establishes $\sum_{k=1}^n u_k(S) \le O((\log\log\log n)^2)\,\E[\OPT_{\cU_S}]$, which is $O(1)\,\E[\OPT_{\cU_S}]$ whenever $n$ is bounded by an absolute constant; thus we may assume throughout that $n \ge 64$.} For any subset $U \subseteq S$, we write $X(U) := |\{i \in [n] : r_i \in U\}| \sim \mathrm{Bin}(n, |U|/n)$ for the number of requests of $R_n \sim \cU_S^n$ landing in $U$. As in Section~\ref{sec:mst-expected-opt}, $T = \MST(S, d)$ denotes a minimum spanning tree of $(S, d)$ and $\wt(T)$ its weight.

The proof has three steps. Section~\ref{subsec:pow2} reduces the problem to the case where $n$ is a power of two (\Cref{lem:pow2-red}). Section~\ref{subsec:two-stage} splits $\tilde{u}_k(S)$ into an intra-cluster term and an inter-cluster term (\Cref{lem:two-stage-decomp}), states the bounds on the two terms (\Cref{lem:stage1-bound,lem:stage2-peeling}), and derives from them the per-step bound (\Cref{lem:per-step-uk}) and Theorem~\ref{thm:main-o1}. Sections~\ref{subsec:stage1-net} and~\ref{subsec:stage2} prove the two bounds.

\subsection{Reduction to Powers of Two}
\label{subsec:pow2}

When considering a sub-instance on a server multiset $S' \subseteq X$ of size $n'$, we write $\OPT_{\cD}(S') := \OPT(R_{n'}, S')$ for the random variable equal to the offline optimal matching cost on $n'$ requests $R_{n'} \sim \cD^{n'}$, with expected value $\E[\OPT_{\cD}(S')] := \E_{R_{n'} \sim \cD^{n'}}[\OPT(R_{n'}, S')]$ (keeping the shorthand $\OPT_{\cD} := \OPT_{\cD}(S)$ when the server multiset is $S$).
To partition $S$ into equal-sized clusters without leftover remainder servers, we first show by a black-box reduction that it suffices to prove $\sum_{k=1}^{n'} u_k(S') \le O(1)\,\E[\OPT_{\cU_{S'}}(S')]$ when the server multiset size $n' = 2^\ell$ is a power of two
.

\begin{lemma}[Reduction to Power-of-Two Server Sizes]
\label{lem:pow2-red}
Suppose there exists a constant $C > 0$ such that $\sum_{k=1}^{n'} u_k(S') \le C\,\E[\OPT_{\cU_{S'}}(S')]$ for every server multiset $S' \subseteq X$ whose size $n'$ is a power of two. Then for every server multiset $S \subseteq X$ of size $n \ge 1$, we have
\[
    \sum_{k=1}^n u_k(S) \;\le\; (4C + 2)\,\E[\OPT_{\cU_S}(S)].
\]
\end{lemma}

\begin{proof}
Let $n' := 2^{\lfloor \log_2 n \rfloor}$, so $\frac{1}{2}n < n' \le n$. If $n' = n$, the claim holds immediately with $C \le 4C + 2$, so assume $n' < n$.

The proof considers an intermediate server multiset $S'$ of size $n'$ drawn from $\cU_{n'}(S)$. In particular, we can draw $S_k \sim \cU_k(S)$ by first drawing $S' \sim \cU_{n'}(S)$ and then drawing $S_k \sim \cU_k(S')$; by the triangle inequality~\ref{prop:w1-triangle}, we have $W_1(\cU_{S_k}, \cU_S) \le W_1(\cU_{S_k}, \cU_{S'}) + W_1(\cU_{S'}, \cU_S)$. Hence, for $1\le k\le n'$, 
\begin{align*}
    u_k(S) & =\E_{S_k\sim\cU_{k}(S)}[W_1(\cU_{S_k},\cU_S)] \\
    & \le  \E_{S'\sim \cU_{n'}(S)}[\E_{S_k\sim\cU_k(S')}[W_1(\cU_{S_k},\cU_{S'})\mid S']] + \E_{S'\sim \cU_{n'}(S)}[\E_{S_k\sim\cU_k(S')}[W_1(\cU_{S'},\cU_{S})\mid S']] \\
    & = \E_{S'}[u_k(S')]+u_{n'}(S).
\end{align*}
Subset monotonicity (\Cref{lem:replacement-mono}\ref{prop:mono}) gives $u_{k}(S)\le u_{n'}(S)$ for $k>n'$. Combining both inequalities yields
\begin{equation}\label{eq:pow-of-2:reduce}
    \sum_{k=1}^n u_k(S) \le \E_{S'\sim \cU_{n'}(S)}\left[\sum_{k=1}^{n'}u_k(S')\right] +n\, u_{n'}(S) \le C\,\E_{S' \sim \cU_{n'}(S)}\!\Bigl[\E[\OPT_{\cU_{S'}}(S')]\Bigr] + n\, u_{n'}(S).
\end{equation}

\paragraph{Subsampling bound on $u_{n'}$.} 
Split $R_n = (r_1, \dots, r_n) \sim \cU_S^n$ into $R_{n'} := (r_1, \dots, r_{n'}) \sim \cU_S^{n'}$ and $R'' := (r_{n'+1}, \dots, r_n) \sim \cU_S^{n - n'}$. Condition on $R_{n'}$. By the replacement lemma (\Cref{lem:replacement-mono}\ref{prop:replace}), \[
    u_{n'}(S)\le \tilde{u}_{n'}(S)=\E_{R_{n'}\sim \cI_{n'}(S)}\bigl[W_1(\cU_{R_{n'}}, \cU_S)\bigr].
\]
Now consider excess and deficit 
\begin{align*}
    \E_{R''}[\cU_{R_n}\mid R_{n'}]-\cU_S & = \E_{R''}\left[\frac{n'}{n}\cU_{R_{n'}} + \frac{n-n'}{n}\cU_{R''} \;
    \middle|\; R_{n'}\right]-\cU_S \\
    & = \frac{n'}{n}\cU_{R_{n'}}+\frac{n-n'}{n}\cU_S-\cU_S \\
    & = \frac{n'}{n}(\cU_{R_{n'}}-\cU_S).
\end{align*}
By shortcut property \ref{prop:w1-shortcut}, $W_1\bigl(\E_{R''}[\cU_{R_n} \mid R_{n'}],\, \cU_S\bigr) = \frac{n'}{n}\,W_1(\cU_{R_{n'}}, \cU_S)$. By Jensen's inequality \ref{prop:w1-convex}, $W_1\bigl(\E_{R''}[\cU_{R_n} \mid R_{n'}],\, \cU_S\bigr) \le \E_{R''}[W_1(\cU_{R_n},\cU_S)\mid R_{n'}]$. Both give $\E_{R''}[W_1(\cU_{R_n},\cU_S)\mid R_{n'}] \ge \frac{n'}{n}\,W_1(\cU_{R_{n'}}, \cU_S)$, so 
\begin{align*}
    \E_{R_{n'}\sim \cI_{n'}(S)}\bigl[W_1(\cU_{R_{n'}}, \cU_S)\bigr] & \le \frac{n}{n'} \E_{R_{n'}\sim \cI_{n'}(S)}\bigl[ \E_{R''}[W_1(\cU_{R_n},\cU_S)\mid R_{n'}] \bigr] \\
    & = \frac{n}{n'}\E_{R_{n}\sim\cU_S^{n}}[W_1(\cU_{R_n},\cU_S)].
\end{align*}
Combining them all with \ref{prop:w1-matching} yields 
\begin{equation}\label{eq:pow-of-2:u_n}
    u_{n'}(S) \le \tilde{u}_{n'}(S) \le \frac{n}{n'}\E_{R_{n}\sim\cU_{S}^{n}}[W_1(\cU_{R_n},\cU_S)] =\frac{1}{n'}\E[\OPT_{\cU_S}(S)].
\end{equation}

\paragraph{Comparing $\E_{S' \sim \cU_{n'}(S)}\!\Bigl[\E[\OPT_{\cU_{S'}}(S')]\Bigr]$ to $\E[\OPT_{\cU_S}(S)]$.} For any $n'$-subset $S' \subseteq S$,
\begin{align*}
    \E[\OPT_{\cU_{S'}}(S')] & \le 2\,\E[\OPT_{\cU_S}(S')] \tag{\Cref{lem:opt-comparison}} \\
    & = 2n'\,\E_{R_{n'} \sim \cU_S^{n'}}[W_1(\cU_{R_{n'}}, \cU_{S'})] \tag*{\ref{prop:w1-matching}} \\
    & \le 2n'(\tilde{u}_{n'}(S) + W_1(\cU_S, \cU_{S'})). \tag*{\ref{prop:w1-triangle}}
\end{align*}
Thus, using \eqref{eq:pow-of-2:u_n},
\begin{equation}\label{eq:pow-of-2:opt-compare}
    \E_{S'\sim \cU_{n'}(S)}\!\Bigl[\E[\OPT_{\cU_{S'}}(S')]\Bigr] \;\le\; 2n'\bigl(\tilde{u}_{n'}(S) + u_{n'}(S)\bigr) \;\le\; 4\,\E[\OPT_{\cU_S}(S)].
\end{equation}

Therefore, combining \eqref{eq:pow-of-2:reduce}, \eqref{eq:pow-of-2:u_n}, and \eqref{eq:pow-of-2:opt-compare} completes the proof.
\end{proof}

\subsection{Two-Stage Decomposition and Proof of Theorem~\ref{thm:main-o1}}
\label{subsec:two-stage}

Because of \Cref{lem:pow2-red}\old{ (together with~\cite[Theorem~1.1]{GGPW19} for $n < 64$)}, we may assume that \old{$n  \ge 64$}\new{$n$} is a power of two. Any power-of-two cluster size $m \in [n]$ then divides $n$ evenly, partitioning $S$ into $B = n/m$ disjoint clusters $C_1, \dots, C_B \subseteq S$ of exact size $m$ so that uniform draws from $S$ induce a uniform distribution across the $B$ clusters. For a divisor $m$ of $n$, we call a partition of $S$ into $B := n/m$ pairwise disjoint clusters $C_1, \dots, C_B \subseteq S$ of size $m$ each an \emph{$m$-clustering} of $S$; whenever an $m$-clustering is fixed, $B$ and $C_1, \dots, C_B$ refer to it.

\begin{definition}[Cluster-Smoothed Distributions]
\label{def:cluster-smoothed}
Fix an $m$-clustering of $S$. The \emph{cluster-smoothed distribution} of any non-empty subset $A \subseteq [B]$ of cluster indices is
\[
    \widetilde{\cU}_A \;:=\; \frac{1}{|A|} \sum_{a \in A} \cU_{C_a},
\]
which spreads weight $1/|A|$ uniformly across each cluster $C_a$ for each $a \in A$. Similarly, for a sequence of $k \ge 1$ cluster indices $\hat{A}_k = (a_1, \dots, a_k) \in [B]^k$, we define $\widetilde{\cU}_{\hat{A}_k} := \frac{1}{k}\sum_{i=1}^k \cU_{C_{a_i}}$. Notice that when $A = [B]$, we have $\widetilde{\cU}_{[B]} = \cU_S$.
\end{definition}

As outlined in Section~\ref{subsec:overview-q1}, we split $u_k(S) \le \tilde{u}_k(S)$ into intra-cluster smoothing (Stage~1) and inter-cluster transport (Stage~2).

\begin{lemma}[Two-Stage Clustered Decomposition]
\label{lem:two-stage-decomp}
For every $n$-point metric space $(S, d)$, every integer $k \in \{1, \dots, n\}$, and every $m$-clustering of $S$, we have
\[
    u_k(S) \;\le\; \tilde{u}_k(S) \;\le\; \underbrace{\frac{1}{n}\sum_{a=1}^B \frac{1}{m}\sum_{u, v \in C_a} d(u, v)}_{\text{\normalfont\textbf{Stage 1} (Intra-Cluster Smoothing)}}
    \;+\; \underbrace{\E_{\hat{A}_k \sim \cI_k([B])}\bigl[W_1(\widetilde{\cU}_{\hat{A}_k}, \cU_S)\bigr]}_{\text{\normalfont\textbf{Stage 2} (Inter-Cluster Transport)}}.
\]
\end{lemma}
\begin{proof}
By the Replacement Lemma (Lemma~\ref{lem:replacement-mono}\ref{prop:replace}), we have $u_k(S) \le \tilde{u}_k(S) = \E_{\hat{S}_k}[W_1(\cU_{\hat{S}_k}, \cU_S)]$ for $\hat{S}_k = (\hat{s}_1, \dots, \hat{s}_k) \sim \cI_k(S)$. Because $C_1, \dots, C_B$ partition $S$ into $B$ clusters of equal size $m$, drawing each $\hat{s}_i \sim \cU_S$ is equivalent to first drawing a random cluster index $a_i \in [B]$ uniformly and then drawing $\hat{s}_i \sim \cU_{C_{a_i}}$ inside $C_{a_i}$. Let $\hat{A}_k = (a_1, \dots, a_k) \sim \cI_k([B])$. By the triangle inequality~\ref{prop:w1-triangle} along the chain $\cU_{\hat{S}_k} \to \widetilde{\cU}_{\hat{A}_k} \to \cU_S$, we have
\[
    \tilde{u}_k(S)
    \;\le\; \E\bigl[W_1(\cU_{\hat{S}_k}, \widetilde{\cU}_{\hat{A}_k})\bigr] + \E\bigl[W_1(\widetilde{\cU}_{\hat{A}_k}, \cU_S)\bigr].
\]
Spreading each point mass $\delta_{\hat{s}_i}$ uniformly over $C_{a_i}$ costs $W_1(\delta_{\hat{s}_i}, \cU_{C_{a_i}}) = \frac{1}{m}\sum_{v \in C_{a_i}} d(\hat{s}_i, v)$. By convexity~\ref{prop:w1-convex} and averaging over $\hat{s}_i \sim \cU_S$, the first term is at most $\frac{1}{n}\sum_{a=1}^B \frac{1}{m}\sum_{u, v \in C_a} d(u, v)$.
\end{proof}

We now state our bounds for Stage~1 (proved in Section~\ref{subsec:stage1-net}) and Stage~2 (proved in Section~\ref{subsec:stage2}), and show how they imply our per-step bound on $u_k(S)$ and Theorem~\ref{thm:main-o1}.

\begin{lemma}[Stage 1 Bound: Intra-Cluster Smoothing]
\label{lem:stage1-bound}
For every $n$-point metric space $(S, d)$ and every divisor $m \in \{1, \dots, n\}$ of $n$, there exists an $m$-clustering of $S$ such that
\[
    \frac{1}{n}\sum_{a=1}^B \frac{1}{m}\sum_{u, v \in C_a} d(u, v)
    \;\le\; \frac{2e\,m}{n}\,\E[\OPT_{\cU_S}].
\]
\end{lemma}

\begin{lemma}[Stage 2 Bound via Multiplicity Peeling]
\label{lem:stage2-peeling}
Let $(S, d)$ be an $n$-point metric space whose size $n \ge 2$ is a power of two. For any power of two $m\in [2,n]$, set $B:=n/m$ and let $k\in [B/2]$.

For $\hat{A}_k = (a_1, \dots, a_k) \sim \cI_k([B])$, let $M_a := |\{i \in [k] : a_i = a\}|$ be the multiplicity of cluster $a \in [B]$, and let $K_j := |\{a \in [B] : M_a \ge j\}|$ for $j \in \{1, \dots, k\}$. Then,
for every $m$-clustering of $S$, we have
    \[
        \E_{\hat{A}_k \sim \cI_k([B])}\bigl[W_1(\widetilde{\cU}_{\hat{A}_k}, \cU_S)\bigr]
        \;\le\; \frac{4}{k\sqrt{m}}\,\E[\OPT_{\cU_S}]
        \;+\; \sum_{j=2}^k \E\left[\frac{K_j}{k}\,\tilde{u}_{K_j}(S)\right].
    \]
\end{lemma}

Combining \Cref{lem:two-stage-decomp} with \Cref{lem:stage1-bound,lem:stage2-peeling} yields a polynomial per-step bound on $\tilde{u}_k(S)$ and $u_k(S)$.

Unlike the general-metric bounds of~\cite[Lemmas~3.1 and~3.8]{GGPW19} (see \Cref{subsec:overview-q1}), \Cref{lem:per-step-uk} achieves the exponent $2/3 < 1$ on $k$ without any logarithmic distortion factor, so its sum over $k \in [n]$ converges directly to $O(1)$. 

\begin{lemma}[Per-Step Transport Bound on General Metrics]
\label{lem:per-step-uk}
Let $(S,d)$ be an $n$-point metric space whose size $n\ge64$ is a power of two. Then, for every positive integer $k\le n/64$,
\begin{equation}
\label{eq:ind-uk}
    \tilde{u}_k(S)
    \le \frac{32}{n^{1/3}k^{2/3}}\,\E[\OPT_{\cU_S}].
\end{equation}
Consequently, for every $k\in[n]$,
\[
    u_k(S)
    \le \frac{512}{n^{1/3}k^{2/3}}\,\E[\OPT_{\cU_S}].
\]
\end{lemma}

\begin{proof}
Fix a positive integer $k\le n/64$. Let $m$ be the largest power of two at most $\frac14(n/k)^{2/3}$, namely
\[
    m:=2^{\lfloor \frac23\log_2(n/k)\rfloor-2},
    \qquad B:=n/m.
\]
Then $4\le m\le n$, $m$ divides $n$, and
\(
    \frac18\left(\frac nk\right)^{2/3}
    <m\le
    \frac14\left(\frac nk\right)^{2/3}.
\)
In particular,
\(
    \frac{k}{B}=\frac{km}{n}
    \le\frac14\left(\frac kn\right)^{1/3}
    \le\frac1{16},
\)
so $k\le B/2$.
\paragraph{Deriving a recurrence relation.}
Choose the $m$-clustering supplied by \Cref{lem:stage1-bound}.
Combining \Cref{lem:two-stage-decomp,lem:stage1-bound,lem:stage2-peeling} gives
\[
    \tilde{u}_k(S)
    \le
    \left(\frac{2em}{n}+\frac{4}{k\sqrt{m}}\right)
    \E[\OPT_{\cU_S}]
    +\sum_{j=2}^k
    \E\left[\frac{K_j}{k}\tilde{u}_{K_j}(S)\right],
\]
where $K_j$ is the number of clusters drawn at least $j$ times.
Our bounds on $m$ imply
\[
    \frac{2em}{n}+\frac{4}{k\sqrt{m}}
    \le
    \frac{e/2+8\sqrt{2}}{n^{1/3}k^{2/3}}
    \le
    \frac{16}{n^{1/3}k^{2/3}}.
\]
Hence,
\begin{equation}
\label{eq:uk-step}
    \tilde{u}_k(S)
    \le
    \frac{16}{n^{1/3}k^{2/3}}\,\E[\OPT_{\cU_S}]
    +\sum_{j=2}^k
    \E\left[\frac{K_j}{k}\tilde{u}_{K_j}(S)\right].
\end{equation}

For each $j\in[k]$, counting the draws in clusters of multiplicity
at least $j$ gives $jK_j\le k$.
Moreover, a union bound over the $\binom{k}{j}$ sets of $j$ draws gives
\[
    \frac{\E[K_j]}{k}
    \le
    \frac{B}{k}\binom{k}{j}B^{-j}
    \le
    \frac1{j!}\left(\frac{k}{B}\right)^{j-1}
    \le
    \frac1{j!\,16^{j-1}}.
\]
By H\"older's inequality with exponents $3$ and $3/2$,
\begin{align}
    \sum_{j=2}^{\infty}
    \left(\frac1{j!\,16^{j-1}}\right)^{1/3}
    &\le
    \left(\sum_{j=2}^{\infty}\frac1{j!}\right)^{1/3}
    \left(\sum_{j=2}^{\infty}\frac1{4^{j-1}}\right)^{2/3}\nonumber\\
    &=\left(\frac{e-2}{9}\right)^{1/3}
    <\left(\frac19\right)^{1/3}
    <\frac12. \label{eq:convergent-series}
\end{align}

\paragraph{Solving the recurrence relation.}
We now prove \eqref{eq:ind-uk} by strong induction on $k\in[n/64]$.
For $k=1$, the sum in \eqref{eq:uk-step} is empty, so the claim follows.
For $k\ge2$, assume the claim holds for every positive integer $r<k$.
Since $K_j\le\lfloor k/j\rfloor<k$ for every $j\ge2$,
the induction hypothesis gives
\[
    \frac{K_j}{k}\tilde{u}_{K_j}(S)
    \le \frac{32}{n^{1/3}{K_j}^{2/3}}\left(\frac{K_j}{k}\right) \E[\OPT_{\cU_S}] =
    \frac{32}{n^{1/3}k^{2/3}}
    \left(\frac{K_j}{k}\right)^{1/3}
    \E[\OPT_{\cU_S}],
\]
where both sides are zero when $K_j=0$.
Thus, although $K_j<k$, weighting layer $j$ by $K_j/k$ leaves only the damping factor $(K_j/k)^{1/3}$, and these factors sum to less than $1/2$ in expectation (see \eqref{eq:convergent-series}).
By Jensen's inequality for the concave function $x\mapsto x^{1/3}$,
\begin{align*}
    \sum_{j=2}^k
    \E\left[\frac{K_j}{k}\tilde{u}_{K_j}(S)\right]
    &\le
    \frac{32}{n^{1/3}k^{2/3}}
    \sum_{j=2}^k
    \left(\frac{\E[K_j]}{k}\right)^{1/3}
    \E[\OPT_{\cU_S}]\\
    &\le
    \frac{16}{n^{1/3}k^{2/3}}\,\E[\OPT_{\cU_S}].
\end{align*}
Substituting into \eqref{eq:uk-step} proves \eqref{eq:ind-uk}.

Finally, for $k\le n/64$, the replacement lemma gives
$u_k(S)\le\tilde{u}_k(S)$, so the claimed bound on $u_k(S)$ follows.
For $k>n/64$, subset monotonicity gives
\[
    u_k(S)
    \le u_{n/64}(S)
    \le \frac{32\cdot64^{2/3}}{n}\,\E[\OPT_{\cU_S}]
    =\frac{512}{n}\,\E[\OPT_{\cU_S}]
    \le\frac{512}{n^{1/3}k^{2/3}}\,\E[\OPT_{\cU_S}].
    \qedhere
\]
\end{proof}

\begin{proof}[Proof of Theorem~\ref{thm:main-o1}]
For every \new{power of two} $n < 64$, we already have $\sum_{k=1}^n u_k(S) \le O(1)\,\E[\OPT_{\cU_S}]$ by~\cite[Theorem~1.1]{GGPW19}\new{, whose bound $O((\log\log\log n)^2)\,\E[\OPT_{\cU_S}]$ is $O(1)\,\E[\OPT_{\cU_S}]$ for bounded $n$}. For every power of two $n \ge 64$, using \Cref{lem:per-step-uk} and $k^{-2/3} \le \int_{k-1}^k x^{-2/3}\,dx$ gives
\[
    \sum_{k=1}^n u_k(S) \;\le\; \left(\frac{512}{n^{1/3}}\int_0^n x^{-2/3}\,dx\right)\E[\OPT_{\cU_S}] \;=\; 1536\,\E[\OPT_{\cU_S}].
\]
Thus $\sum_{k=1}^n u_k(S) \le O(1)\,\E[\OPT_{\cU_S}]$ whenever $n$ is a power of two, and applying \Cref{lem:pow2-red} together with Lemmas~\ref{lem:opt-comparison}~and~\ref{lem:ggpw-structure}(b)  yields $\E[\FB(\cD)] \le O(1)\,\E[\OPT_{\cD}]$.
\end{proof}
\subsection{Cluster Construction and Proof of the Stage 1 Bound}
\label{subsec:stage1-net}

We now prove \Cref{lem:stage1-bound} by ordering $S$ along a depth-first pre-order traversal of a Minimum Spanning Tree $T = \MST(S, d)$ (as in the ``twice-around-the-tree'' tour for the metric Traveling Salesperson Problem~\cite{RSL77}) and slicing the tour into $B = n/m$ contiguous blocks of size $m$ (as in iterated tour partitioning for capacitated routing~\cite{HK85}); see \Cref{alg:cluster-partition}. Because the $B$ cluster segments are disjoint portions of a tour of length at most $2\,\wt(T)$, bounding Stage~1 reduces to bounding $\wt(T)$ against $\E[\OPT_{\cU_S}]$, and \Cref{thm:mst-sharp-bound} gives $\wt(T) \le e\,\E[\OPT_{\cU_S}]$.

\begin{algorithm}[t]
\caption{Cluster Partition via MST Tour Slicing}
\label{alg:cluster-partition}
\vspace{0.3em}
\noindent\textbf{Input:} Metric space $(S, d)$ of size $n$ and divisor $m \in \{1, \dots, n\}$ of $n$.\\
\noindent\textbf{Output:} An $m$-clustering $C_1, \dots, C_B$ of $S$.
\begin{enumerate}[leftmargin=2em,label=\arabic*:,itemsep=0.25em,topsep=0.3em]
    \item \textbf{MST Tour of Servers:} Construct a Minimum Spanning Tree $T = \MST(S, d)$, and let $(s'_1, \dots, s'_n)$ list the servers in the order in which a depth-first traversal of $T$ first visits them (the pre-order).
    \item \textbf{Contiguous Cluster Slicing:} Partition $(s'_1, \dots, s'_n)$ into $B := n/m$ contiguous blocks $C_a := \{s'_{(a-1)m+1}, \dots, s'_{am}\}$ (for $a = 1, \dots, B$) of size $m$.
    \item \textbf{Return} the $m$-clustering $C_1, \dots, C_B$.
\end{enumerate}
\vspace{0.1em}
\end{algorithm}

\begin{proof}[Proof of Lemma~\ref{lem:stage1-bound}]
Let $C_1, \dots, C_B$ be the $m$-clustering produced by Algorithm~\ref{alg:cluster-partition} from $T = \MST(S, d)$. The depth-first traversal walks every edge of $T$ twice, and it reaches $s'_{i+1}$ from $s'_i$ along stretches that are pairwise disjoint for different $i$, so the triangle inequality gives $\sum_{i=1}^{n-1} d(s'_i, s'_{i+1}) \le 2\,\wt(T)$. For each cluster $C_a = \{s'_{(a-1)m+1}, \dots, s'_{am}\}$ and any $u = s'_p$ and $v = s'_q$ in $C_a$ ($p \le q$), the triangle inequality along the tour gives
\[
    d(u, v) \;\le\; \sum_{i=p}^{q-1} d(s'_i, s'_{i+1}) \;\le\; \sum_{i=(a-1)m+1}^{am-1} d(s'_i, s'_{i+1}),
\]
so $\frac{1}{m}\sum_{u, v \in C_a} d(u, v) \le m\sum_{i=(a-1)m+1}^{am-1} d(s'_i, s'_{i+1})$. Since the index blocks of $C_1, \dots, C_B$ are pairwise disjoint and $B m = n$, summing over $a = 1, \dots, B$ and applying \Cref{thm:mst-sharp-bound} (with $C_n < e$) yields
\[
    \frac{1}{n}\sum_{a=1}^B \frac{1}{m}\sum_{u, v \in C_a} d(u, v)
    \;\le\; \frac{m}{n}\sum_{i=1}^{n-1} d(s'_i, s'_{i+1})
    \;\le\; \frac{2m}{n}\,\wt(T)
    \;\le\; \frac{2e\,m}{n}\,\E[\OPT_{\cU_S}].
    \qedhere
\]
\end{proof}

\subsection{Proof of the Stage 2 Bound}
\label{subsec:stage2}


We now prove \Cref{lem:stage2-peeling}. Throughout this subsection, we fix an $m$-clustering $C_1, \dots, C_B$ of $S$ with $B \ge 2$ (for $B = 1$ the lemma is vacuous, since $[B/2]$ is empty), write $\cC := [B]$ for the set of cluster indices, and let $X_a := X(C_a)$ be the number of requests of $R_n \sim \cU_S^n$ landing in cluster $C_a$. The central quantity is the expected cost of $r$ distinct uniformly random clusters:
\[
    w_r \;:=\; \E_{A \sim \cU_r(\cC)}\bigl[W_1(\widetilde{\cU}_A, \cU_S)\bigr] \quad \text{for } r \in [B], \qquad \text{and} \qquad w_0 := 0.
\]
The proof rests on three lemmas. The first peels the $k$ drawn clusters into multiplicity layers, each of which is a uniformly random set of distinct clusters. The second bounds every layer by the cost of the same number of independent server draws; we use it for the collision layers $j \ge 2$. The third is a stronger bound for the first layer, which comes from the $\Omega(\sqrt{m})$ fluctuations of the cluster loads.

\begin{lemma}[Peeling into Multiplicity Layers]
\label{lem:peeling}
For every integer $k \ge 1$ and $\hat{A}_k \sim \cI_k(\cC)$, with $K_j$ the number of clusters drawn at least $j$ times as in \Cref{lem:stage2-peeling},
\[
    \E\bigl[W_1(\widetilde{\cU}_{\hat{A}_k}, \cU_S)\bigr] \;\le\; \sum_{j=1}^k \E\left[\frac{K_j}{k}\,w_{K_j}\right].
\]
\end{lemma}

\begin{lemma}[Collision Layers]
\label{lem:collision-layers}
For every $r \in [B]$, we have $w_r \le \tilde{u}_r(S)$.
\end{lemma}

\begin{lemma}[First Layer]
\label{lem:first-layer}
For every positive integer $r \le B/2$, we have $w_r \le \frac{4}{r\sqrt{m}}\,\E[\OPT_{\cU_S}]$.
\end{lemma}

\begin{proof}[Proof of \Cref{lem:stage2-peeling}]
For $k \in [B/2]$, the first layer satisfies $1 \le K_1 \le k \le B/2$, so \Cref{lem:first-layer} gives
\[
    \frac{K_1}{k}\,w_{K_1} \;\le\; \frac{K_1}{k} \cdot \frac{4}{K_1\sqrt{m}}\,\E[\OPT_{\cU_S}] \;=\; \frac{4}{k\sqrt{m}}\,\E[\OPT_{\cU_S}].
\]
For $j \ge 2$, \Cref{lem:collision-layers} gives $w_{K_j} \le \tilde{u}_{K_j}(S)$. Substituting these bounds into \Cref{lem:peeling} proves the lemma.
\end{proof}

\begin{proof}[Proof of \Cref{lem:peeling}]
Recall that $M_a = |\{i \in [k] : a_i = a\}|$ is the multiplicity of cluster $a \in \cC$ in $\hat{A}_k = (a_1, \dots, a_k)$ and that $K_j = |\{a \in \cC : M_a \ge j\}|$. For each $j \in [k]$, define the $j$-th \emph{multiplicity layer} $A^{(j)} := \{a \in \cC : M_a \ge j\}$, so that $|A^{(j)}| = K_j$. For example, the draws $(1, 1, 2)$ have layers $A^{(1)} = \{1, 2\}$ and $A^{(2)} = \{1\}$. Cluster $a$ belongs to exactly the layers $1, \dots, M_a$, so
\[
    \widetilde{\cU}_{\hat{A}_k} \;=\; \sum_{a \in \cC} \frac{M_a}{k}\,\cU_{C_a} \;=\; \sum_{j=1}^k \frac{1}{k}\sum_{a \in A^{(j)}} \cU_{C_a} \;=\; \sum_{j \in [k] :\, K_j \ge 1} \frac{K_j}{k}\,\widetilde{\cU}_{A^{(j)}},
\]
which is a convex combination because $\sum_{j=1}^k K_j = \sum_{a \in \cC} M_a = k$. By Jensen's inequality~\ref{prop:w1-convex},
\[
    W_1(\widetilde{\cU}_{\hat{A}_k}, \cU_S) \;\le\; \sum_{j \in [k] :\, K_j \ge 1} \frac{K_j}{k}\,W_1(\widetilde{\cU}_{A^{(j)}}, \cU_S).
\]
Because the draws $a_1, \ldots, a_k$ are uniform over $\cC$, the distribution of the multiplicity  $(M_1, \ldots, M_B)$ is invariant under permutations of $\cC$. Hence, conditioned on $K_j = r$ for any $r \ge 1$, the layer $A^{(j)}$ is a uniformly random $r$-element subset of $\cC$, and therefore
\[
    \E\bigl[W_1(\widetilde{\cU}_{A^{(j)}}, \cU_S) \bigm| K_j = r\bigr] \;=\; w_r.
\]
Therefore, taking expectations in the sum over nonempty layers and using $w_0=0$, we obtain
\[
    \E\bigl[W_1(\widetilde{\cU}_{\hat{A}_k},\cU_S)\bigr] \;\le\; \sum_{j=1}^k \E\left[\frac{K_j}{k}\,w_{K_j}\right]. \qedhere
\]
\end{proof}

\begin{proof}[Proof of \Cref{lem:collision-layers}]
The replacement lemma (\Cref{lem:replacement-mono}\ref{prop:replace}) gives the first inequality below. Since drawing $\hat{S}_r \sim \cI_r(S)$ is equivalent to drawing $\hat{A}_r \sim \cI_r(\cC)$ and then one uniform server from each chosen cluster, $\E[\cU_{\hat{S}_r} \mid \hat{A}_r] = \widetilde{\cU}_{\hat{A}_r}$, and Jensen's inequality~\ref{prop:w1-convex} gives
\[
    w_r \;\le\; \E_{\hat{A}_r \sim \cI_r(\cC)}\bigl[W_1(\widetilde{\cU}_{\hat{A}_r}, \cU_S)\bigr]
    \;\le\; \E_{\hat{S}_r \sim \cI_r(S)}\bigl[W_1(\cU_{\hat{S}_r}, \cU_S)\bigr]
    \;=\; \tilde{u}_r(S). \qedhere
\]
\end{proof}

It remains to prove \Cref{lem:first-layer}. As in the balls-and-bins argument of~\cite{GGPW19}, we compare a uniformly random set of $r$ clusters with the $r$ most loaded clusters.

\begin{definition}[Top-$r$ Cluster Load Excess]
\label{def:cluster-excess}
For each $r \in [B]$, the \emph{top-$r$ cluster load} is
\[
    L_r \;:=\; \max_{A \subseteq [B],\, |A| = r} \sum_{a \in A} X_a,
\]
the total number of requests in the $r$ most loaded clusters, and the \emph{expected top-$r$ cluster load excess} over their capacity $r m$ is $\Delta_r := \E[L_r] - r m$.
\end{definition}

\begin{lemma}[Clustered Balls-and-Bins Lemma]
\label{lem:clustered-bb}
For every $r \in [B-1]$ with $\Delta_r > 0$ (Definition~\ref{def:cluster-excess}), we have
\[
    w_r \;\le\; \frac{1}{\Delta_r}\,\E[\OPT_{\cU_S}].
\]
\end{lemma}
\begin{proof}

\emph{Step 1: symmetry.} Let $A^* \subseteq \cC$ be the set of the $r$ clusters with the largest loads $X_a$, where ties between equal loads are broken by a uniformly random permutation $\sigma$ of $\cC$ that is independent of $R_n$ (among clusters of equal load, the one placed earlier by $\sigma$ is preferred); thus $L_r = \sum_{a \in A^*} X_a$. Since all clusters have size $m$, each request is equally likely to land in any cluster, so the joint distribution of the loads $(X_a)_{a \in \cC}$ and $\sigma$ is unchanged when the cluster indices are permuted. Permuting the indices by (coupling) $\pi$ replaces $A^*$ by $\pi(A^*)$ and leaves $L_r$ unchanged, so $(\pi(A^*), L_r)$ has the same distribution as $(A^*, L_r)$ for every permutation $\pi$ of $\cC$. Consequently: (i) every $r$-subset $A \subseteq \cC$ is equally likely to be $A^*$, so $\Pr(A^* = A) = \binom{B}{r}^{-1}$ and $A^* \sim \cU_r(\cC)$; (ii) the conditional distribution of $L_r$ given $A^* = A$ does not depend on $A$, that is, the maximum value $L_r$ is independent of the maximizer $A^*$; and (iii) every permutation that maps $A$ onto itself preserves the event $\{A^* = A\}$, so conditioned on $A^* = A$, the loads of the clusters in $A$ are exchangeable, and so are the loads of the clusters outside $A$. In particular, $\E[X_a \mid A^* = A]$ is the same for all $a \in A$, and the same for all $a \in \cC \setminus A$.

\emph{Step 2: the conditional request distribution.} Fix an $r$-subset $A \subseteq \cC$. By (iii) and (ii), $\E[X_a \mid A^* = A] = \frac{1}{r}\E[L_r \mid A^* = A] = \frac{1}{r}\E[L_r]$ for $a \in A$, and likewise $\E[X_a \mid A^* = A] = \frac{1}{B-r}(n - \E[L_r])$ for $a \in \cC \setminus A$. Given the loads and $\sigma$ (which determine whether $A^* = A$), the requests in each cluster are uniform within it, so $\E[\cU_{R_n} \mid A^* = A] = \sum_{a \in \cC} \frac{1}{n}\E[X_a \mid A^* = A]\,\cU_{C_a}$; that is,
\[
    \E[\cU_{R_n} \mid A^* = A] \;=\; \frac{\E[L_r]}{n}\,\widetilde{\cU}_A + \Bigl(1 - \frac{\E[L_r]}{n}\Bigr)\widetilde{\cU}_{\cC \setminus A},
    \qquad \text{whereas} \qquad
    \cU_S \;=\; \frac{rm}{n}\,\widetilde{\cU}_A + \Bigl(1 - \frac{rm}{n}\Bigr)\widetilde{\cU}_{\cC \setminus A}.
\]
Subtracting gives
\[
    \E[\cU_{R_n} \mid A^* = A] - \cU_S
    \;=\; \frac{\E[L_r] - rm}{n}\,(\widetilde{\cU}_A-\widetilde{\cU}_{\cC \setminus A}) 
    \;=\; \frac{\Delta_r}{n}\,(\widetilde{\cU}_A - \widetilde{\cU}_{\cC \setminus A}).
\]
In words, conditioning on $A$ being the set of the most loaded clusters tilts the request distribution from $\cU_S$ toward $\widetilde{\cU}_A$.

\emph{Step 3: the cost.} By conditional probability, we have $\cU_S=\frac{r}{B}\widetilde{\cU}_A+\frac{B-r}{B}\widetilde{\cU}_{\cC\setminus A}$. That is, we can transport $\widetilde{\cU}_A$ to $\cU_S$ by keeping $\frac{r}{B}$ mass in place and moving the remaining $\frac{B-r}{B}$ mass from $\widetilde{\cU}_A$ to $\widetilde{\cU}_{\cC\setminus A}$, yielding 
\begin{equation}\label{eq:clustered-bb1}
    W_1(\widetilde{\cU}_A,{\cU}_{S}) \le \frac{B-r}{B} W_1(\widetilde{\cU}_A,\widetilde{\cU}_{\cC\setminus A}).
\end{equation}
By Step~2, $\E[\cU_{R_n} \mid A^* = A]$ has net excess $\frac{\Delta_r}{n}\widetilde{\cU}_A$ and net deficit $\frac{\Delta_r}{n}\widetilde{\cU}_{\cC \setminus A}$ relative to $\cU_S$, so the shortcut property~\ref{prop:w1-shortcut} gives $W_1(\E[\cU_{R_n} \mid A^* = A], \cU_S) = \frac{\Delta_r}{n}\,W_1(\widetilde{\cU}_A, \widetilde{\cU}_{\cC \setminus A})$. Together with equivalence to a minimum bipartite matching \ref{prop:w1-matching} and Jensen's inequality~\ref{prop:w1-convex},
\begin{align*}
    \E[\OPT_{\cU_S} \mid A^* = A] & \;=\; n\,\E[W_1(\cU_{R_n}, \cU_S) \mid A^* = A] \tag*{\ref{prop:w1-matching}} \\
    & \;\ge\; n\,W_1\bigl(\E[\cU_{R_n} \mid A^* = A], \cU_S\bigr) \tag*{\ref{prop:w1-convex}} \\
    & \;=\; \Delta_r\,W_1(\widetilde{\cU}_A, \widetilde{\cU}_{\cC\setminus A}) \tag*{\ref{prop:w1-shortcut}} \\
    & \;\ge\; \frac{B}{B-r}\Delta_r\,W_1(\widetilde{\cU}_A, \cU_S). \tag{from \eqref{eq:clustered-bb1}}
\end{align*}
Therefore, since $\frac{B}{B-r} \ge 1$, taking expectations over $A^* \sim \cU_r(\cC)$ gives
\begin{align*}
    \E[\OPT_{\cU_S}] & \;=\; \sum_{A \subseteq \cC :\, |A| = r} \Pr(A^* = A)\,\E[\OPT_{\cU_S} \mid A^* = A] \\
    & \;\ge\; \Delta_r\,\E_{A \sim \cU_r(\cC)}\bigl[W_1(\widetilde{\cU}_A, \cU_S)\bigr] \;=\; \Delta_r\cdot w_r. \qedhere
\end{align*}
\end{proof}

Next, we lower-bound the excess load $\Delta_r$. Because the loads $(X_1, \dots, X_B)$ are negatively correlated, realizations with fewer than $r$ overloaded clusters ($X_a > m$) force any $r$-subset $A$ to include non-overloaded clusters; we handle both cases pointwise via the conservation law $\sum_{a=1}^B (X_a - m) = 0$, which equates the excess on $A$ to the deficit outside $A$. 

\begin{lemma}[Top-$r$ Cluster Load Excess Bound]
\label{lem:top-r-load}
For every positive integer $r \le B/2$, we have
\[
    \Delta_r \;=\; \E[L_r] - r m \;\ge\; \frac{1}{4}\, r \sqrt{m}.
\]
\end{lemma}
\begin{proof}
For any realization of $R_n$, let $P:=\{a\in [B]:X_a> m\}$ be the set of overloaded clusters. 
Since $\sum_{a=1}^B (X_a-m)=0$, the total excess $E:=\sum_{a\in P} (X_a-m)=\sum_{a\notin P} (m-X_a)=\frac{1}{2}\sum_{a=1}^B|X_a-m|$. We consider two cases depending on whether there is nonpositive contribution from a cluster to the sum $L_r-rm$.
\begin{itemize}
    \item $|P|\ge r$: By an averaging argument, we have $L_r-rm\ge \frac{r}{|P|}\sum_{a\in P} (X_a-m)\ge \frac{rE}{B}$.
    \item $|P|<r$: In this case, $L_r-rm=E-\min_{A\subseteq [B]\setminus P:|A|=r-|P|} \sum_{a\in A}(m-X_a)$. By an averaging argument, the negative term is at most $\frac{r-|P|}{B-|P|}E$. We thus have $L_r-rm\ge \frac{B-r}{B-|P|}E\ge \frac{rE}{B}$ (since $r\le B/2$).
\end{itemize}

In both cases, $L_r-rm \ge \frac{rE}{B} = \frac{r}{2B}\sum_{a=1}^B |X_a - m|$. Since $X_a \sim \mathrm{Bin}(n, 1/B)$ and $B\ge 2$, we can apply \Cref{lem:bin-mad}, yielding $\Delta_r = \E[L_r]-rm \ge \frac{r}{2}\E[|X_1-m|] \ge \frac{r}{4}\sqrt{m}$.
\end{proof}

\begin{proof}[Proof of \Cref{lem:first-layer}]
Let $r \le B/2$ be a positive integer. Then $r \le B - 1$, and $\Delta_r \ge \frac{1}{4}r\sqrt{m} > 0$ by \Cref{lem:top-r-load}, so \Cref{lem:clustered-bb} gives $w_r \le \frac{1}{\Delta_r}\,\E[\OPT_{\cU_S}] \le \frac{4}{r\sqrt{m}}\,\E[\OPT_{\cU_S}]$.
\end{proof}

\section{Unknown Arrival Distributions (Question 2)}
\label{sec:unknown}

We now study the setting in which the arrival distribution $\cD$ over $X$ is \emph{unknown} to the online algorithm upfront. Throughout this section, we fix a finite metric space $(X, d)$, a server multiset $S \subseteq X$ of size $n \ge 1$, and an arrival distribution $\cD$ over $X$, and the requests $R_n = (r_1, \dots, r_n)$ are drawn i.i.d.\ from $\cD$. For $t \in [n]$, we write $\hat{\cD}_t := \frac{1}{t}\sum_{i=1}^t \delta_{r_i}$ for the empirical distribution of the first $t$ requests, and we let $\hat{\cD}_0 := 0$ be the zero measure.

\subsection{The \texorpdfstring{\EFB{}}{Empirical-Fair-Bias} Algorithm and the Uniform-Remaining-Servers Invariant}
\label{subsec:efb}

\EFB{} (\Cref{alg:efb}) uses the sampling rule of \FB{} with $\cD$ replaced by an empirical distribution. Two design choices make this work. First, at step $t \in [n]$, the empirical distribution is formed \emph{after} observing $r_t$ and includes it: this \textbf{inclusive empirical distribution} $\hat{\cD}_t\,=\,\frac{1}{t}\sum_{i=1}^t \delta_{r_i}$ satisfies $\hat{\cD}_t(r_t) \ge 1/t > 0$ and depends only on the unordered multiset $\{r_1, \dots, r_t\}$. Second, the optimal transport plan $\pi^{(t)} \in \Pi(\cU_{S_k}, \hat{\cD}_t)$ is \emph{canonical}: it is chosen by a fixed deterministic rule (for instance, the lexicographically first optimal basic solution) that depends only on $(S_k, \hat{\cD}_t)$ and not on the order in which $r_1, \dots, r_t$ arrived.

\begin{algorithm}[t]
\caption{\EFB{} (for an Unknown Arrival Distribution $\cD$)}
\label{alg:efb}
\vspace{0.3em}
\noindent\textbf{Input:} Server multiset $S \subseteq X$ of size $n$ in metric space $(X, d)$.
\begin{enumerate}[leftmargin=2em,label=\arabic*.,itemsep=0.25em,topsep=0.3em]
    \item Initialize free server set $S_n \gets S$.
    \item \textbf{for} $t = 1, 2, \dots, n$ \textbf{do}
    \begin{enumerate}[leftmargin=1.8em,label=(\alph*),itemsep=0.2em,topsep=0.15em]
        \item Let $k \gets n - t + 1$.
        \item Observe arriving request $r_t \in X$, and form the inclusive empirical distribution \\ $\hat{\cD}_t \gets \frac{1}{t}\sum_{i=1}^t \delta_{r_i}$.
        \item Compute a canonical optimal transport plan $\pi^{(t)} \in \Pi(\cU_{S_k}, \hat{\cD}_t)$ achieving $W_1(\cU_{S_k}, \hat{\cD}_t)$, chosen as a deterministic function of $(S_k, \hat{\cD}_t)$ (independent of the arrival order of $r_1, \dots, r_t$).
        \item Sample a server $s_t \in S_k$ with probability $\Pr(s_t = s \mid S_k, \hat{\cD}_t, r_t) = \frac{\pi^{(t)}(s, r_t)}{\hat{\cD}_t(r_t)}$.
        \item Match $r_t$ to $s_t$, and update $S_{k-1} \gets S_k \setminus \{s_t\}$.
    \end{enumerate}
\end{enumerate}
\vspace{0.1em}
\end{algorithm}

As explained in \Cref{subsec:overview-q2}, including $r_t$ in $\hat{\cD}_t$ and breaking ties symmetrically across $\{r_1, \dots, r_t\}$ ensures that the conditional arrival probability $\Pr(r_t = x \mid \hat{\cD}_t) = \hat{\cD}_t(x)$ cancels the denominator in Step~2(d) of \Cref{alg:efb} at every step. We first record this fact.

\begin{lemma}[The Last Request Given the Empirical Distribution]
\label{lem:last-request}
For every $t \in [n]$ and every $x \in X$, we have $\Pr(r_t = x \mid \hat{\cD}_t) = \hat{\cD}_t(x)$.
\end{lemma}
\begin{proof}
Since $r_1, \dots, r_t$ are i.i.d., every ordering of the multiset $\{r_1, \dots, r_t\}$ is equally likely. Hence, given the multiset (equivalently, given $\hat{\cD}_t$), the last request $r_t$ is a uniformly random one of the $t$ requests, and $\Pr(r_t = x \mid \hat{\cD}_t) = \frac{1}{t}|\{i \in [t] : r_i = x\}| = \hat{\cD}_t(x)$.
\end{proof}

\begin{lemma}[Uniform-Remaining-Servers Invariant for \EFB{}]
\label{lem:exchangeable-structure}
For every step $t \in [n]$ (with $k = n - t + 1$), the following hold under \EFB{} (\Cref{alg:efb}):
\begin{enumerate}[leftmargin=2em,label=(\alph*)]
    \item \emph{(Uniformity and independence)} At the start of step $t$, the remaining free server set $S_k$ is uniformly distributed over all $k$-subsets of $S$ ($S_k \sim \cU_k(S)$) and is \emph{statistically independent} of the past empirical distribution $\hat{\cD}_{t-1}$ (and hence jointly independent of $(\hat{\cD}_t, r_t)$).
    \item \emph{(Expected step cost)} The expected matching cost incurred at step $t$ is
    \[
        \E\bigl[d(s_t, r_t)\bigr] \;=\; \E_{S_k,\hat{\cD}_t}\bigl[W_1(\cU_{S_k}, \hat{\cD}_t)\bigr] \;=\; \E_{S_k \sim \cU_k(S),\; (r_1, \dots, r_t) \sim \cD^t}\bigl[W_1(\cU_{S_k}, \hat{\cD}_t)\bigr],
    \] where the server subset and the request sequence in the final expectation are sampled independently.
\end{enumerate}
\end{lemma}
\begin{proof}

We prove (a) by induction on $t$. At $t = 1$, both $S_n = S$ and $\hat{\cD}_0$ are deterministic, so (a) holds. Suppose that (a) holds at the start of step $t$.

\emph{$S_k$ is independent of $(\hat{\cD}_t, r_t)$.} The request $r_t$ is independent of the history before step $t$, and hence of $(S_k, \hat{\cD}_{t-1})$. Together with the induction hypothesis, $S_k$, $\hat{\cD}_{t-1}$, and $r_t$ are mutually independent. Since $\hat{\cD}_t = \frac{t-1}{t}\hat{\cD}_{t-1} + \frac{1}{t}\delta_{r_t}$ is determined by $(\hat{\cD}_{t-1}, r_t)$, the set $S_k$ is independent of $(\hat{\cD}_t, r_t)$.

\emph{The matched server is uniform.} Condition on $(S_k, \hat{\cD}_t)$, which fixes the canonical plan $\pi^{(t)}$. By the preceding paragraph, additionally conditioning on $S_k$ does not change the distribution of $r_t$ given $\hat{\cD}_t$, so \Cref{lem:last-request} gives $\Pr(r_t = x \mid S_k, \hat{\cD}_t) = \hat{\cD}_t(x)$. Since $\pi^{(t)}(s, x) = 0$ whenever $\hat{\cD}_t(x) = 0$, for every $s \in S_k$,
\[
    \Pr\bigl(s_t = s \bigm| S_k, \hat{\cD}_t\bigr)
    \;=\; \sum_{x :\, \hat{\cD}_t(x) > 0} \hat{\cD}_t(x) \cdot \frac{\pi^{(t)}(s, x)}{\hat{\cD}_t(x)}
    \;=\; \sum_{x} \pi^{(t)}(s, x) \;=\; \cU_{S_k}(s) \;=\; \frac{1}{k}.
\]
Thus, given $(S_k, \hat{\cD}_t)$, the set $S_{k-1} = S_k \setminus \{s_t\}$ is obtained by deleting a uniformly random server from $S_k$, and this conditional distribution does not depend on $\hat{\cD}_t$. Since $S_k \sim \cU_k(S)$ is independent of $\hat{\cD}_t$, it follows that $S_{k-1} \sim \cU_{k-1}(S)$ and that $S_{k-1}$ is independent of $\hat{\cD}_t$, which is (a) at the start of step $t + 1$.

\emph{The expected cost.} In the same way,
\begin{align*}
    \E\bigl[d(s_t, r_t) \bigm| S_k, \hat{\cD}_t\bigr]
    & \;=\; \sum_{x :\, \hat{\cD}_t(x) > 0} \hat{\cD}_t(x) \sum_{s \in S_k} \frac{\pi^{(t)}(s, x)}{\hat{\cD}_t(x)}\, d(s, x) \\
    & \;=\; \sum_{s \in S_k,\, x} \pi^{(t)}(s, x)\, d(s, x) \;=\; W_1(\cU_{S_k}, \hat{\cD}_t),
\end{align*}
and taking expectations over the independent pair $(S_k, \hat{\cD}_t)$ proves (b).
\end{proof}

\subsection{3-Hop Time-Reversal Coupling and Proof of Theorem~\ref{thm:main-unknown}}
\label{subsec:time-reversal}

It remains to bound $\E[W_1(\cU_{S_k}, \hat{\cD}_t)]$. A triangle inequality through $\cD$ would require bounding $\E[W_1(\hat{\cD}_t, \cD)]$, which does not depend on $S$ and can be large. Instead, we route through the servers along the chain
\[
    \cU_{S_k} \;\longrightarrow\; \cU_S \;\longrightarrow\; \cU_{\tilde{S}_t} \;\longrightarrow\; \cU_{\tilde{R}_t},
\]
where $\tilde{R}_t$ is a uniformly random $t$-subset of all $n$ requests (which has the same distribution as the first $t$ requests) and $\tilde{S}_t$ is the set of servers matched to $\tilde{R}_t$ by an offline optimal matching. The first two hops cost $u_k(S)$ and $u_t(S)$, because $S_k$ and $\tilde{S}_t$ are uniformly random subsets of $S$, and the last hop costs $\frac{1}{n}\E[\OPT_{\cD}]$ in expectation, because each matched pair is used with probability $t/n$.

\begin{remark}[Time-Reversal Coupling]
\label{rem:time-reversal}
Time-reversal coupling refers to how we view the first $t$ arrivals. Instead of generating the requests forward in time, we first draw the whole sequence $R_n \sim \cD^n$, together with an offline optimal matching $\sigma^*$ for it, and only then reveal which $t$ of the $n$ requests arrived first. Since the requests are i.i.d., the multiset of the first $t$ requests has the same distribution as a uniformly random $t$-subset $\tilde{R}_t$ of $R_n$. Looking back from the end of the sequence in this way lets us compare the requests seen so far with the servers that the offline optimum assigns to them, even though that matching depends on requests that have not yet arrived. 
\end{remark}
\begin{lemma}[3-Hop Time-Reversal Coupling]
\label{lem:3hop-coupling}
For every 
step $t \in [n]$ ($k = n - t + 1$), let $S_k \sim \cU_k(S)$ 
be independent of $R_n$. Then
\[
    \E\bigl[W_1(\cU_{S_k}, \hat{\cD}_t)\bigr] \;\le\; u_k(S) + u_t(S) + \frac{1}{n}\,\E[\OPT_{\cD}].
\]
\end{lemma}
\begin{proof}
Because all $n$ requests in $R_n = (r_1, \dots, r_n) \sim \cD^n$ are drawn i.i.d.\ from $\cD$, the distribution of the multiset of the first $t$ requests is identical to the distribution obtained by first drawing all $n$ requests $R_n \sim \cD^n$ and then selecting a uniformly random $t$-element subset of indices $I \sim \cU_t([n])$ independently of $(S_k, R_n)$. Let $\tilde{R}_t := \{r_i : i \in I\}$ be the multiset of $t$ sampled requests; then $\cU_{\tilde{R}_t} = \frac{1}{t}\sum_{i \in I} \delta_{r_i}$ has the exact same distribution as $\hat{\cD}_t$.

For each realization of $R_n$, let $\sigma^* : [n] \to [n]$ be an offline optimal one-to-one matching minimizing $\sum_{i=1}^n d(r_i, s_{\sigma^*(i)}) = \OPT(R_n, S)$, and let $\tilde{S}_t := \{s_{\sigma^*(i)} : i \in I\} \subseteq S$ be the subset of $t$ servers matched by $\sigma^*$ to the requests in $\tilde{R}_t$. We observe two facts:
\begin{enumerate}[leftmargin=2em,label=(\arabic*)]
    \item Conditioned on $R_n$, since $\sigma^* : [n] \to [n]$ is a one-to-one correspondence and $I$ is a uniformly random $t$-subset of $[n]$, the matched subset $\tilde{S}_t$ is a uniformly random $t$-subset of $S$; that is, $\tilde{S}_t \sim \cU_t(S)$. Thus $\E[W_1(\cU_S, \cU_{\tilde{S}_t})] = u_t(S)$.
    \item Moving weight $1/t$ from each server $s_{\sigma^*(i)} \in \tilde{S}_t$ to its partner $r_i$ ($i \in I$) gives a valid transport plan from $\cU_{\tilde{S}_t}$ to $\cU_{\tilde{R}_t}$ of cost $\frac{1}{t}\sum_{i \in I} d(s_{\sigma^*(i)}, r_i)$. Since each index $i \in [n]$ belongs to the random $t$-subset $I$ with probability $t/n$, taking expectation over $I$ and then $R_n$ gives
    \[
        \E\bigl[W_1(\cU_{\tilde{S}_t}, \cU_{\tilde{R}_t})\bigr]
        \;\le\; \E_{R_n \sim \cD^n}\!\left[\frac{1}{t}\sum_{i=1}^n \frac{t}{n}\, d(s_{\sigma^*(i)}, r_i)\right]
        \;=\; \frac{1}{n}\,\E[\OPT_{\cD}].
    \]
\end{enumerate}
Applying the triangle inequality~\ref{prop:w1-triangle} along the 3-hop chain $\cU_{S_k} \to \cU_S \to \cU_{\tilde{S}_t} \to \cU_{\tilde{R}_t}$ gives
\begin{align*}
    \E\bigl[W_1(\cU_{S_k}, \hat{\cD}_t)\bigr] \;=\; \E\bigl[W_1(\cU_{S_k}, \cU_{\tilde{R}_t})\bigr]
    &\;\le\; \E\bigl[W_1(\cU_{S_k}, \cU_S)\bigr] + \E\bigl[W_1(\cU_S, \cU_{\tilde{S}_t})\bigr] + \E\bigl[W_1(\cU_{\tilde{S}_t}, \cU_{\tilde{R}_t})\bigr] \\
    &\;\le\; u_k(S) + u_t(S) + \frac{1}{n}\,\E[\OPT_{\cD}].
    \qedhere
\end{align*}
\end{proof}

\begin{proof}[Proof of \Cref{thm:main-unknown}]
Combining \Cref{lem:exchangeable-structure,lem:3hop-coupling} gives
\begin{align*}
    \E[\EFB(\cD)]
    &\;=\; \sum_{t=1}^n \E\bigl[W_1(\cU_{S_{n-t+1}}, \hat{\cD}_t)\bigr] \\
    &\;\le\; \sum_{t=1}^n \left(u_{n-t+1}(S) + u_t(S) + \frac{1}{n}\,\E[\OPT_{\cD}]\right) \\
    &\;=\; 2 \sum_{k=1}^n u_k(S) + \E[\OPT_{\cD}].
\end{align*}
Therefore, by \Cref{lem:ggpw-structure}, $\E[\EFB(\cD)] \,\le\, 2 \, \E[\FB(\cU_S)] + \E[\OPT_{\cD}].$

In particular, suppose that $\E[\FB(\cU_S)] \le \rho\,\E[\OPT_{\cU_S}]$ for some $\rho \ge 1$, that is, \FB{}$(\cU_S)$ is $\rho$-competitive on $(S, d)$. Since $\E[\OPT_{\cU_S}] \le 2\,\E[\OPT_{\cD}]$ by \Cref{lem:opt-comparison}, we obtain
\[
    \E[\EFB(\cD)] \;\le\; (4\rho + 1)\,\E[\OPT_{\cD}]. \qedhere
\]
\end{proof}


\section{Open Questions}
\label{sec:open-questions}
Our analysis shows that both \textsc{Fair-Bias} and \textsc{Empirical-Fair-Bias} are $O(1)$-competitive, closing the asymptotic gap for online metric matching with i.i.d.\ arrivals on general metrics. However, the competitive ratios obtained from our analysis exceed 10000, which is far from the known upper bound of $9$ on tree metrics under known arrival distributions \cite{GGPW19}. It is an interesting question to determine the  exact optimal competitive ratio.

Unlike \cite{GGPW19}, our analysis does not rely on tree embeddings. Together with the tight asymptotic bounds for adversarial ($\Theta(n)$)~\cite{KP93,KMV94} and random-order ($\Theta(\log n)$)~\cite{Rag16} arrival models, both of which are obtained without tree embeddings, this leaves the oblivious-adversary model as the only remaining arrival model for which the asymptotic competitive ratio is not yet settled. The best known bounds for this model are $O(\log^2 n)$~\cite{BBGN14} and $\Omega(\log n)$~\cite{MNP06}, with the extra logarithmic factor arising from tree embedding. It would be particularly interesting to close this gap, perhaps by developing an analysis that avoids tree embeddings altogether.


\section*{AI Disclosure} We used AI models at several stages of this work. Specifically, we asked Gemini 4 Carbon whether clustering could help for \emph{unknown} arrival distributions. Instead, the model pointed out that clustering can give $O(1)$-competitiveness for \emph{known} distributions (i.e. \Cref{thm:main-o1}), and that one can simply use the \textsc{Empirical-Fair-Bias} for \emph{unknown} distributions (i.e. \Cref{thm:main-unknown}). We then simplified the proofs further, including suggesting the reduction to power-of-two (\Cref{subsec:pow2}) and connecting the expected optimum to cut relaxation LP for MST (\Cref{sec:mst-expected-opt}). 

Additionally, ChatGPT-6 Astra helped draft the discussion of related work. Claude Fable 5.2 and Opus 5.5 assisted in revising the manuscript: checking proofs, restructuring and simplifying their presentation, unifying notation, and suggesting references. All AI suggestions were reviewed by the authors, who modified, rewrote, or discarded many of them, and all AI-assisted proofs were checked in detail and in many cases rewritten from scratch. The authors verified the correctness and originality of all content and every reference, and take full responsibility for the paper.

\bibliographystyle{alpha}
\bibliography{references}

@article{BBGN14,
  author  = {Bansal, Nikhil and Buchbinder, Niv and Gupta, Anupam and Naor, Joseph (Seffi)},
  title   = {A randomized {$O(\log^2 k)$}-competitive algorithm for metric bipartite matching},
  journal = {Algorithmica},
  volume  = {68},
  number  = {2},
  pages   = {390--403},
  year    = {2014}
  
}

@article{BK13,
  author  = {Berend, Daniel and Kontorovich, Aryeh},
  title   = {A sharp estimate of the binomial mean absolute deviation with applications},
  journal = {Statistics \& Probability Letters},
  volume  = {83},
  number  = {4},
  pages   = {1254--1259},
  year    = {2013}
}

@article{FRT04,
  author  = {Fakcharoenphol, Jittat and Rao, Satish and Talwar, Kunal},
  title   = {A tight bound on approximating arbitrary metrics by tree metrics},
  journal = {Journal of Computer and System Sciences},
  volume  = {69},
  number  = {3},
  pages   = {485--497},
  year    = {2004}
}

@inproceedings{GGPW19,
  author    = {Gupta, Anupam and Guruganesh, Guru and Peng, Binghui and Wajc, David},
  title     = {Stochastic online metric matching},
  booktitle = {Proceedings of the 46th International Colloquium on Automata, Languages, and Programming (ICALP)},
  pages     = {67:1--67:14},
  year      = {2019},
  note = {(Full version: arXiv:1904.09284)}
}

@inproceedings{GL12,
  author    = {Gupta, Anupam and Lewi, Kevin},
  title     = {The online metric matching problem for doubling metrics},
  booktitle = {Proceedings of the 39th International Colloquium on Automata, Languages, and Programming (ICALP)},
  pages     = {424--435},
  year      = {2012}
}

@article{HK85,
  author  = {Haimovich, Mordechai and {Rinnooy Kan}, Alexander H. G.},
  title   = {Bounds and heuristics for capacitated routing problems},
  journal = {Mathematics of Operations Research},
  volume  = {10},
  number  = {4},
  pages   = {527--542},
  year    = {1985}
}

@article{KP93,
  author  = {Kalyanasundaram, Bala and Pruhs, Kirk},
  title   = {Online weighted matching},
  journal = {Journal of Algorithms},
  volume  = {14},
  number  = {3},
  pages   = {478--488},
  year    = {1993}
}

@article{Kan22,
  author  = {Kanoria, Yashodhan},
  title   = {Dynamic spatial matching},
  journal = {The Annals of Applied Probability},
  volume  = {35},
  number  = {5},
  pages   = {3086--3118},
  year    = {2025}
}

@article{KMV94,
  author  = {Khuller, Samir and Mitchell, Stephen G. and Vazirani, Vijay V.},
  title   = {On-line algorithms for weighted bipartite matching and stable marriages},
  journal = {Theoretical Computer Science},
  volume  = {127},
  number  = {2},
  pages   = {255--267},
  year    = {1994}
}

@inproceedings{LVY26,
  author    = {Li, Yingxi and Vitercik, Ellen and Yang, Mingwei},
  title     = {Smoothed analysis of online metric matching with a single sample: Beyond metric distortion},
  booktitle = {Proceedings of the 17th Innovations in Theoretical Computer Science Conference (ITCS)},
  pages     = {94:1--94:23},
  year      = {2026}
}

@inproceedings{MNP06,
  author    = {Meyerson, Adam and Nanavati, Amit and Poplawski, Laura J.},
  title     = {Randomized online algorithms for minimum metric bipartite matching},
  booktitle = {Proceedings of the 17th Annual ACM-SIAM Symposium on Discrete Algorithms (SODA)},
  pages     = {954--959},
  year      = {2006}
}

@article{PS21,
  author  = {Peserico, Enoch and Scquizzato, Michele},
  title   = {Matching on the line admits no $o(\sqrt{\log n})$-competitive algorithm},
  journal = {ACM Transactions on Algorithms},
  volume  = {19},
  number  = {3},
  pages   = {28:1--28:4},
  year    = {2023},
  doi     = {10.1145/3594873}
}

@article{PC19,
  author  = {Peyr{\'e}, Gabriel and Cuturi, Marco},
  title   = {Computational optimal transport: With applications to data science},
  journal = {Foundations and Trends in Machine Learning},
  volume  = {11},
  number  = {5--6},
  pages   = {355--607},
  year    = {2019}
}

@inproceedings{Rag16,
  author    = {Raghvendra, Sharath},
  title     = {A robust and optimal online algorithm for minimum metric bipartite matching},
  booktitle = {Approximation, Randomization, and Combinatorial Optimization. Algorithms and Techniques (APPROX/RANDOM 2016)},
  pages     = {18:1--18:16},
  year      = {2016}
}

@inproceedings{Rag18,
  author    = {Raghvendra, Sharath},
  title     = {Optimal analysis of an online algorithm for the bipartite matching problem on a line},
  booktitle = {Proceedings of the 34th International Symposium on Computational Geometry (SoCG)},
  pages     = {67:1--67:14},
  year      = {2018}
}

@article{RSL77,
  author  = {Rosenkrantz, Daniel J. and Stearns, Richard E. and {Lewis II}, Philip M.},
  title   = {An analysis of several heuristics for the traveling salesman problem},
  journal = {SIAM Journal on Computing},
  volume  = {6},
  number  = {3},
  pages   = {563--581},
  year    = {1977}
}

@inproceedings{SYY24,
  author    = {Saberi, Amin and Yang, Mingwei and Yu, Sophie H.},
  title     = {Stochastic online metric matching: Adversarial is no harder than stochastic},
  booktitle = {Proceedings of the 20th International Conference on Web and Internet Economics (WINE)},
  pages     = {102--118},
  year      = {2024}
}

@book{Vil09,
  author    = {Villani, C{\'e}dric},
  title     = {Optimal Transport: Old and New},
  series    = {Grundlehren der mathematischen Wissenschaften},
  volume    = {338},
  publisher = {Springer},
  year      = {2009}
}

@article{YY26,
  author  = {Yang, Mingwei and Yu, Sophie H.},
  title   = {Online metric matching: Beyond the worst case},
  journal = {Operations Research},
  volume  = {74},
  number  = {1},
  pages   = {130--140},
  year    = {2026}
}

@inproceedings{KN04,
  author    = {Koutsoupias, Elias and Nanavati, Akash},
  title     = {The online matching problem on a line},
  booktitle = {Proceedings of the 1st International Workshop on Approximation and Online Algorithms (WAOA 2003)},
  series    = {Lecture Notes in Computer Science},
  volume    = {2909},
  pages     = {179--191},
  publisher = {Springer},
  year      = {2004},
  doi       = {10.1007/978-3-540-24592-6_14}
}

@article{FHK05,
  author  = {Fuchs, Bernhard and Hochst{\"a}ttler, Winfried and Kern, Walter},
  title   = {Online matching on a line},
  journal = {Theoretical Computer Science},
  volume  = {332},
  number  = {1--3},
  pages   = {251--264},
  year    = {2005},
  doi     = {10.1016/j.tcs.2004.10.028}
}

@article{ABNPS19,
  author  = {Antoniadis, Antonios and Barcelo, Neal and Nugent, Michael and Pruhs, Kirk and Scquizzato, Michele},
  title   = {A {$o(n)$}-competitive deterministic algorithm for online matching on a line},
  journal = {Algorithmica},
  volume  = {81},
  number  = {7},
  pages   = {2917--2933},
  year    = {2019},
  doi     = {10.1007/s00453-019-00565-w}
}

@inproceedings{NR17,
  author    = {Nayyar, Krati and Raghvendra, Sharath},
  title     = {An input sensitive online algorithm for the metric bipartite matching problem},
  booktitle = {Proceedings of the 58th IEEE Annual Symposium on Foundations of Computer Science (FOCS)},
  pages     = {505--515},
  year      = {2017},
  doi       = {10.1109/FOCS.2017.53}
}

@inproceedings{AFT18,
  author    = {Antoniadis, Antonios and Fischer, Carsten and T{\"o}nnis, Andreas},
  title     = {A collection of lower bounds for online matching on the line},
  booktitle = {Proceedings of the 13th Latin American Symposium on Theoretical Informatics (LATIN)},
  series    = {Lecture Notes in Computer Science},
  volume    = {10807},
  pages     = {52--65},
  publisher = {Springer},
  year      = {2018},
  doi       = {10.1007/978-3-319-77404-6_5}
}

@misc{GRV26,
  author        = {Gor, Parth and Roy, Sourya and Varadarajan, Kasturi},
  title         = {Near-optimal online metric matching on {$\Delta$}-ary {HST}},
  year          = {2026},
  eprint        = {2609.25292},
  archivePrefix = {arXiv},
  primaryClass  = {cs.DS},
  url           = {https://arxiv.org/abs/2609.25292},
  note          = {arXiv:2609.25292. To appear in \emph{Proceedings of SOFSEM 2027}}
}

@article{GK19,
  author  = {Gairing, Martin and Klimm, Max},
  title   = {Greedy metric minimum online matchings with random arrivals},
  journal = {Operations Research Letters},
  volume  = {47},
  number  = {2},
  pages   = {88--91},
  year    = {2019},
  doi     = {10.1016/j.orl.2019.01.002}
}

@misc{Har26,
  author        = {Harada, Tsubasa},
  title         = {Online metric matching with unknown {i.i.d.} arrivals},
  year          = {2026},
  eprint        = {2611.04275},
  archivePrefix = {arXiv},
  primaryClass  = {cs.DS},
  url           = {https://arxiv.org/abs/2611.04275},
  note          = {arXiv preprint arXiv:2611.04275}
}

@inproceedings{BFP23,
  author    = {Balkanski, Eric and Faenza, Yuri and P{\'e}rivier, No{\'e}mie},
  title     = {The power of greedy for online minimum cost matching on the line},
  booktitle = {Proceedings of the 24th ACM Conference on Economics and Computation (EC)},
  pages     = {185--205},
  year      = {2023},
  doi       = {10.1145/3580507.3597794}
}

@article{DSX22,
  author  = {Duppala, Sharmila V. S. and Sankararaman, Karthik A. and Xu, Pan},
  title   = {Online minimum matching with uniform metric and random arrivals},
  journal = {Operations Research Letters},
  volume  = {50},
  number  = {1},
  pages   = {45--49},
  year    = {2022},
  doi     = {10.1016/j.orl.2021.12.005}
}

@article{ACEPS23,
  author  = {Antoniadis, Antonios and Coester, Christian and Eli{\'a}{\v{s}}, Marek and Polak, Adam and Simon, Bertrand},
  title   = {Online metric algorithms with untrusted predictions},
  journal = {ACM Transactions on Algorithms},
  volume  = {19},
  number  = {2},
  pages   = {19:1--19:34},
  year    = {2023},
  doi     = {10.1145/3582689}
}

@inproceedings{SV26,
  author    = {Shin, Yongho and Vajanopath, Phanu},
  title     = {Parsimonious learning-augmented online metric matching},
  booktitle = {Proceedings of the 43rd International Conference on Machine Learning (ICML)},
  series    = {Proceedings of Machine Learning Research},
  volume    = {306},
  pages     = {112669--112688},
  publisher = {PMLR},
  year      = {2026},
  url       = {https://proceedings.mlr.press/v306/shin26e.html}
}

@article{MN25,
  author  = {Megow, Nicole and N{\"o}lke, Lukas},
  title   = {Online metric matching on the line with recourse},
  journal = {Algorithmica},
  volume  = {87},
  number  = {6},
  pages   = {813--841},
  year    = {2025},
  doi     = {10.1007/s00453-025-01299-8}
}

@inproceedings{GKS20,
  author    = {Gupta, Varun and Krishnaswamy, Ravishankar and Sandeep, Sai},
  title     = {{Permutation} strikes back: The power of recourse in online metric matching},
  booktitle = {Approximation, Randomization, and Combinatorial Optimization. Algorithms and Techniques (APPROX/RANDOM 2020)},
  series    = {Leibniz International Proceedings in Informatics},
  volume    = {176},
  pages     = {40:1--40:20},
  year      = {2020},
  doi       = {10.4230/LIPIcs.APPROX/RANDOM.2020.40}
}

@article{DZ91,
  author  = {Diaconis, Persi and Zabell, Sandy},
  title   = {Closed form summation for classical distributions: Variations on a theme of {De Moivre}},
  journal = {Statistical Science},
  volume  = {6},
  number  = {3},
  pages   = {284--302},
  year    = {1991},
  doi     = {10.1214/ss/1177011699}
}

@article{CP08,
  author = {Csaba, B{\'e}la and Pluh{\'a}r, Andr{\'a}s},
  title = {A randomized algorithm for the on-line weighted bipartite matching problem},
  journal = {Journal of Scheduling},
  volume = {11},
  number = {6},
  pages = {449--455},
  year = {2008},
  url = {https://arxiv.org/abs/0705.4673}
}

@article{CN24,
  author = {Csaba, B{\'e}la and {Nagy-Gy{\"o}rgy}, Judit},
  title = {On the advice complexity of online matching on the line},
  journal = {Discrete Mathematics \& Theoretical Computer Science},
  volume = {26},
  number = {3},
  year = {2024},
  doi = {10.46298/dmtcs.14125}
}

@misc{CCM25,
  author = {Canonne, Cl{\'e}ment L. and Chen, Kenny and Mestre, Juli{\'a}n},
  title = {With a little help from my friends: Exploiting probability distribution advice in algorithm design},
  year = {2025},
  eprint = {2505.04949},
  archivePrefix = {arXiv},
  primaryClass = {cs.DS},
  url = {https://arxiv.org/abs/2505.04949},
  note = {To appear in Proceedings of the International Symposium on Algorithms and Computation (ISAAC), 2026}
}

@article{GW95,
  author  = {Michel X. Goemans and David P. Williamson},
  title   = {A General Approximation Technique for Constrained Forest Problems},
  journal = {SIAM Journal on Computing},
  volume  = {24},
  number  = {2},
  pages   = {296--317},
  year    = {1995},
  doi     = {10.1137/S0097539793242618}
}

@article{AKR95,
  author    = {Agrawal, Ajit and Klein, Philip and Ravi, R.},
  title     = {When Trees Collide: An Approximation Algorithm for the Generalized {Steiner} Problem on Networks},
  journal   = {SIAM Journal on Computing},
  volume    = {24},
  number    = {3},
  pages     = {440--456},
  year      = {1995},
  doi       = {10.1137/S0097539792236237}
}

\newpage

\appendix
\appendix
\section{Deferred Proofs from Section~\ref{sec:prelim}}
\label{app:prelim-proofs}

In this appendix, we give short self-contained proofs of Lemmas~\ref{lem:ggpw-structure}, \ref{lem:replacement-mono}, and~\ref{lem:opt-comparison}.

\begin{proof}[Proof of Lemma~\ref{lem:ggpw-structure}]
We prove (a) by induction on $t$. At $t = 1$ (so $k = n$), we have $S_n = S \sim \cU_n(S)$. Suppose $S_k \sim \cU_k(S)$ at the start of step $t$. Conditioned on $S_k$, since $r_t \sim \cD$ is independent of $S_k$, the probability that server $s \in S_k$ is matched at step $t$ is
\[
    \sum_{x : \cD(x) > 0} \cD(x) \cdot \frac{\pi^{(t)}(s, x)}{\cD(x)} \;=\; \sum_{x \in X} \pi^{(t)}(s, x) \;=\; \cU_{S_k}(s) \;=\; \frac{1}{k}.
\]
Because every server in $S_k$ is removed with equal probability $1/k$, removing the matched server leaves a uniformly random $(k-1)$-subset $S_{k-1} \sim \cU_{k-1}(S)$. Moreover, conditioned on $S_k$, the expected step-$t$ distance is $\sum_{x : \cD(x) > 0} \cD(x) \sum_{s \in S_k} \frac{\pi^{(t)}(s, x)}{\cD(x)} d(s, x) = \sum_{s, x} \pi^{(t)}(s, x) d(s, x) = W_1(\cU_{S_k}, \cD)$. Taking expectation over $S_k$ gives (a).

For (b), by the triangle inequality~\ref{prop:w1-triangle}, we have $W_1(\cU_{S_k}, \cD) \le W_1(\cU_{S_k}, \cU_S) + W_1(\cU_S, \cD)$. Summing over $k = 1, \dots, n$ gives $\E[\FB(\cD)] \le \sum_{k=1}^n u_k(S) + n\,W_1(\cU_S, \cD)$. Finally, for $R_n = (r_1, \dots, r_n) \sim \cD^n$, the expected empirical request distribution is $\E[\cU_{R_n}] = \cD$, so Jensen's inequality~\ref{prop:w1-convex} and~\ref{prop:w1-matching} give $n\,W_1(\cU_S, \cD) = n\,W_1(\cU_S, \E[\cU_{R_n}]) \le \E[n\,W_1(\cU_S, \cU_{R_n})] = \E[\OPT_{\cD}]$.
\end{proof}

\begin{proof}[Proof of Lemma~\ref{lem:replacement-mono}]
For (a), generate a random $k$-subset $S_k \sim \cU_k(S)$ in two stages: first draw a random $k'$-subset $S_{k'} \sim \cU_{k'}(S)$, and then draw a random $k$-subset $S_k \sim \cU_k(S_{k'})$ from inside $S_{k'}$. Conditioned on $S_{k'}$, each server in $S_{k'}$ belongs to $S_k$ with probability $k/k'$, so $\E[\cU_{S_k} \mid S_{k'}] = \cU_{S_{k'}}$. By Jensen's inequality~\ref{prop:w1-convex}, we have $W_1(\cU_{S_{k'}}, \cU_S) \le \E[W_1(\cU_{S_k}, \cU_S) \mid S_{k'}]$. Taking expectation over $S_{k'}$ gives $u_{k'}(S) \le u_k(S)$.

For (b), for any multiset $M$ of $k$ indices in $[B]$ with multiplicity vector $\chi_M \in \R^B_{\ge 0}$, the value $P(M) := W_1\bigl(\frac{1}{k}\sum_{a=1}^B \chi_M(a)\mu_a, \nu\bigr)$ is the minimum of a linear program whose equality constraints $\sum_{y \in X} \pi(x, y) = \frac{1}{k}\sum_{a=1}^B \chi_M(a)\mu_a(x)$ are affine in $\chi_M$. Hence~\cite[Lemma~B.1]{GGPW19} applies directly to give $\E_{A \sim \cU_k([B])}[P(A)] \le \E_{\hat{A}_k \sim \cI_k([B])}[P(\hat{A}_k)]$; specializing to $B = n$, $\mu_i = \delta_{s_i}$, and $\nu = \cU_S$ yields $u_k(S) \le \tilde{u}_k(S)$~\cite[Lemma~4.3]{GGPW19}.
\end{proof}

\begin{proof}[Proof of Lemma~\ref{lem:opt-comparison}]
Let $\pi^* \in \Pi(\cU_S, \cD)$ be an optimal transport plan achieving $W_1(\cU_S, \cD)$. Generating $R'_n = (r'_1, \dots, r'_n) \sim \cU_S^n$ and $R_n = (r_1, \dots, r_n) \sim \cD^n$ jointly by drawing $n$ independent pairs $(r'_i, r_i) \sim \pi^*$, the triangle inequality gives $\OPT(R'_n, S) \le \sum_{i=1}^n d(r'_i, r_i) + \OPT(R_n, S)$. Taking expectations and using $n\,W_1(\cU_S, \cD) \le \E[\OPT_{\cD}]$ (from the proof of Lemma~\ref{lem:ggpw-structure}(b)) yields $\E[\OPT_{\cU_S}] \le n\,W_1(\cU_S, \cD) + \E[\OPT_{\cD}] \le 2\,\E[\OPT_{\cD}]$.
\end{proof}

\section{Table of Notation}
\begin{table}[p]
\centering
\caption{Notation.}
\label{tab:notation}
\small
\renewcommand{\arraystretch}{1.1}
\begin{tabular}{@{}l@{\hspace{1.2em}}p{0.68\linewidth}@{}}
\toprule
\textbf{Symbol} & \textbf{Meaning} \\
\midrule
\multicolumn{2}{@{}l}{\textit{Instance and costs}} \\
$(X,d)$ & finite metric space \\
$S$, $n$ & multiset of servers with distinct labels, and $n=|S|$ \\
$\cD$, $R_n$ & arrival distribution, and requests $R_n=(r_1,\dots,r_n)\sim\cD^n$ \\
$S_k$ & set of the $k=n-t+1$ free servers when $r_t$ arrives \\
$\OPT(R_n,S)$, $\OPT_{\cD}$ & minimum perfect-matching cost, and its value for $R_n\sim\cD^n$ \\
\FB{} & algorithm for known $\cD$ (Algorithm~\ref{alg:fb}) \\
\EFB{} & algorithm for unknown $\cD$ (Algorithm~\ref{alg:efb}) \\
$\rho$ & ratio with $\E[\FB(\cU_S)]\le\rho\,\E[\OPT_{\cU_S}]$ \\
\midrule
\multicolumn{2}{@{}l}{\textit{Transport and sampling}} \\
$\delta_x$, $\cU_A$ & point mass at $x$, and uniform measure on a multiset $A$ \\
$\Pi(\mu,\nu)$, $W_1$ & transport plans from $\mu$ to $\nu$, and their minimum cost \\
$\hat{\cD}_t$ & empirical distribution $\frac1t\sum_{i=1}^t\delta_{r_i}$ \\
$\pi^{(t)}$ & optimal plan at step $t$ from $\cU_{S_k}$ to $\cD$ (or to $\hat{\cD}_t$) \\
$\cU_k(A)$, $\cI_k(A)$ & $k$ uniform samples from $A$ without and with replacement \\
$u_k(S)$, $\tilde u_k(S)$ & $\E[W_1(\cU_A,\cU_S)]$ for $A\sim\cU_k(S)$ and $A\sim\cI_k(S)$; $\tilde u_0(S):=0$ \\
\midrule
\multicolumn{2}{@{}l}{\textit{Spanning tree and clusters}} \\
$T$, $\wt(T)$ & minimum spanning tree of $(S,d)$, and its weight \\
$C_n$ & sharp constant $(\frac{n}{n-1})^{n-1}<e$ \\
$m$, $B$, $C_a$ & cluster size, $B=n/m$ clusters, and cluster $a\in[B]$ \\
$\widetilde{\cU}_A$ & cluster-smoothed measure $\frac{1}{|A|}\sum_{a\in A}\cU_{C_a}$ \\
$\hat A_k$, $M_a$ & $k$ i.i.d.\ uniform cluster indices, and multiplicity of $a$ \\
$A^{(j)}$, $K_j$ & multiplicity layer $\{a:M_a\ge j\}$, and $K_j=|A^{(j)}|$ \\
$w_r$ & $\E[W_1(\widetilde{\cU}_A,\cU_S)]$ for $A\sim\cU_r([B])$; $w_0:=0$ \\
$X(U)$, $X_a$ & numbers of requests in $U\subseteq S$ and in $C_a$ under $R_n\sim\cU_S^n$ \\
$L_r$, $\Delta_r$ & load of the $r$ most loaded clusters, and $\Delta_r=\E[L_r]-rm$ \\
\midrule
\multicolumn{2}{@{}l}{\textit{Coupling for unknown arrivals}} \\
$\sigma^*$ & offline optimal matching between $R_n$ and $S$ \\
$\tilde R_t$, $\tilde S_t$ & requests at $t$ random indices, and their servers under $\sigma^*$ \\
\bottomrule
\end{tabular}
\end{table}

\end{document}